\documentclass[namedreferences]{solarphysics}
\usepackage[optionalrh]{spr-sola-addons}
\usepackage{graphicx}
\usepackage{color}           % For color text: \color command
\usepackage{url}             % For breaking URLs easily trough lines
\newcommand{\etal}{{\it et al.}}

\begin{document}

\begin{article}

\begin{opening}

\title{Torsional Oscillations in a Global Solar Dynamo}

\author{P.~\surname{Beaudoin}$^{1}$\sep
        P.~\surname{Charbonneau}$^{1}$\sep
        E.~\surname{Racine}$^{2}$\sep
        P.K.~\surname{Smolarkiewicz}$^{3}$
       }
       
\runningauthor{P. Beaudoin \etal}
\runningtitle{Torsional Oscillations in a Global Solar Dynamo}

   \institute{$^{1}$ D\'epartement de Physique, Universit\'e de Montr\'eal
		     , C.P. 6128, Succ. Centre-Ville, Montr\'eal, Qu\'ebec, H3C 3J7, Canada \\
              $^{2}$ Canadian Space Agency, St-Hubert, Qu\'ebec, Canada \\
              $^{3}$ National Center for Atmospheric Research, Boulder, CO 80307-3000, USA
             }
             
\begin{abstract}
We characterize and analyze rotational torsional oscillations developing
in a large-eddy magnetohydrodynamical simulation of solar
convection (Ghizaru, Charbonneau, and Smolarkiewicz, 
{\it Astrophys. J. Lett.} {\bf 715}, L133 (2010); Racine \etal, {\it Astrophys. J.}
{\bf 735}, 46 (2011)) producing
an axisymmetric large-scale magnetic field undergoing periodic
polarity reversals. Motivated by the many
solar-like features exhibited by these oscillations, we carry out an
analysis of the large-scale zonal dynamics.
We demonstrate that simulated torsional oscillations
are not driven primarily by the periodically-varying large-scale magnetic
torque, as one might have expected, but rather via the magnetic
modulation of angular-momentum transport by the large-scale meridional flow.
This result is confirmed by a straightforward energy analysis.
We also detect
a fairly sharp transition in rotational dynamics taking place
as one moves from the base of the convecting layers to the base of
the thin tachocline-like shear layer formed in the stably stratified
fluid layers immediately below.
We conclude by discussing the implications of our analyses with regards
to the mechanism of amplitude saturation in the global dynamo operating
in the simulation, and speculate on the
possible precursor value of torsional oscillations
for the forecast of solar cycle characteristics.
\end{abstract}
\end{opening}

\section{Introduction}

Differential rotation of the solar surface was first noted in the seventeenth century by Christoph Scheiner on the basis of his extensive sunspot observations. In his 1632 {\it Rosa Ursina}, Scheiner states that sunspots move more slowly the farther away they are from the solar Equator, and even concludes that ``...From this phenomenon is drawn the strongest argument for a fluid surface of the Sun.'' ({\it Rosa Ursina}, p. 559; as cited and translated by \opencite{Mitchell1916}, p. 440). Rediscovered in the mid-nineteenth century by R.C.~Carrington and G.~Sp\"orer and soon thereafter extended to high latitudes by Doppler measurements, solar differential rotation has now been mapped deep into the Sun by helioseismology \cite{CD2002,Howe2009}. From the frequency splitting of acoustic eigenmodes of varying azimuthal orders, it has now been shown that the surface latitudinal differential-rotation pattern, with the solar Equator rotating approximately 30\% faster than the Poles, persists throughout the bulk of the solar convective envelope, down to $r/R\approx 0.71$ (with $R$ the Sun's radius), where it abruptly vanishes across a thin spherical shear layer, known as the tachocline, located immediately beneath the core--envelope interface. The underlying stably stratified core appears to be rotating rigidly (or nearly so) down to $r/R\approx 0.3$, at a rate equal to that of the surface mid-latitudes ({\it e.g.} \opencite{Howe2009}, and references therein).

This internal differential-rotation pattern has remained generally steady since the first helioseismic rotational inversions carried out in the late 1980s; but not {\it exactly} steady.
Rotational torsional oscillations were first noted in surface Doppler measurements \cite{Howard1980}, and later shown by helioseismology to extend all the way to the base of the Sun's convective envelope. The torsional oscillation signal reaches only a few nHz in amplitude (about 0.5\% of the rotational frequency), and peaks at high latitudes and in surface and subsurface layers. The oscillations develop at twice the magnetic-cycle frequency and, at mid- to high latitudes (where the signal is the strongest), retain the same phase at all depths ({\it e.g.} Figure~26 in \opencite{Howe2009}). More elaborate phasing patterns occur with latitudes, with two diverging
``branches'' of faster rotating fluid appearing at mid-latitudes around solar-activity minimum: one migrating all the way down to the Equator in the span of two full activity cycles, the other migrating poleward to cause a marked spin-up of the polar region peaking at around the time of activity maximum (see Figure~25 in \opencite{Howe2009}, and accompanying discussion).

Numerous models have been suggested to explain the observed behavior of solar torsional oscillations, the vast majority relying directly or indirectly on the Lorentz force associated with the Sun's magnetic field. \inlinecite{Howe2009} in Section~9.5 gives a succinct overview of these various theoretical explanations, which turn out to be difficult to confirm or refute on the basis of extant observations.
Torsional oscillations having higher amplitudes near the surface and at high latitudes are certainly to be expected; subjected to a torque of a given magnitude (and of whatever origin), a ring of fluid centered on the solar rotation axis will experience greater angular acceleration if located high up in the envelope since its moment of inertia will be reduced through the lower density; and at a given depth, that same moment will also be smaller at higher latitude because of the shorter moment arm, yielding again greater angular acceleration.
At any rate, both the good phase locking with the magnetic-activity cycle as measured, {\it e.g.} through the sunspot number, and the close tracking of the equatorially migrating band of rotational acceleration with the activity belts, point toward a close dynamical link between torsional oscillations and the cyclically varying large-scale solar magnetic field.
This has in fact remained the favored explanation ever since the discovery of torsional oscillations \cite{Schuessler1981,Yoshimura1981}. Interest in this possible dynamical linkage has in fact recently ramped up, due primarily to the curious observation that the poleward branch of the torsional oscillations, due to appear in the final years of Cycle 23, has failed to show up as anticipated.
Taken together with other peculiar features of the extended activity minimum having followed Cycle 23, this has prompted speculations regarding the possibility that the Sun is about to enter a phase of strongly suppressed magnetic activity, perhaps akin to the Maunder Minimum \cite{Hill2011}.

From a dynamical point of view, the simplest hypothesis would be to assume that torsional oscillations are directly driven by the Lorentz force associated with the cyclic large-scale magnetic component that we associate with the magnetic-activity cycle, acting on the zonal flow as a time-varying perturbation of the global hydrodynamical (HD) balance setting the form of solar internal differential rotation. 
Such a balance would involve Reynolds stresses, Maxwell stresses, and angular-momentum advection by the meridional flow within the convection zone\footnote{Some level of dynamical coupling to the underlying stably stratified radiative core
may also play an important role; on the possible participation of the outer radiative core in setting the angular-momentum balance within the convection zone; see the prescient analysis presented in \inlinecite{Gilman1989}.}. As will become apparent in what follows, the situation may well be far more complex.

In this article, we present an analysis of the dynamics and energetics of torsional oscillations arising in a global magnetohydrodynamical (MHD) simulation of solar convection producing solar-like cycles in its dynamo-generated large-scale magnetic field.
We first (Section~2) give an overview of the simulation itself, together with a description of torsional oscillations arising therein.
We then recast the azimuthal component of the momentum equation in conservative form, which allows the study of azimuthal force balance in terms of fluxes of angular momentum and their temporal variations (Section~3). We also examine the energetics of torsional oscillations, and conclude (Section~4) by elaborating on some consequences of our analysis for dynamo saturation, and for the possible use of torsional oscillations as precursors of cycle-amplitude fluctuations.

\section{Numerical Data}

\subsection{The Global Simulation}

We use numerical data produced by one of the global implicit large-eddy simulations (ILES) of
MHD solar convection of the type presented by \inlinecite{Ghizaru2010},
and \inlinecite{Racine2011}. These remain unique so far in producing an axisymmetric
large-scale magnetic-field component undergoing cyclic polarity reversals on a
multi-decadal timescale, in a manner similar in many ways to what is observed on the Sun.
The underlying mathematical and computational frameworks are described
by \inlinecite{Smolarkiewicz2012}, and represent a MHD generalization of the
well-proven general-purpose geophysical flow simulation code EULAG 
(see \opencite{Prusa2008}, and references therein). A unique feature
of both EULAG and EULAG--MHD is the possibility to delegate all dissipation
to the underlying advection scheme, which
makes it possible to reach a maximally turbulent state at a given grid size,
with stability persisting even when field gradients reach spatial scales commensurate
with the computational cell size.
The interested reader is referred to \inlinecite{Smolarkiewicz2012} for further details on algorithmic implementation.

The MPDATA algorithm (\opencite{Smolarkiewicz1998}; \opencite{Smolarkiewicz2006})
at the core of
both EULAG and EULAG--MHD belongs to the class of advection algorithms
known as non-oscillatory forward-in-time (NFT),
which also includes a number of advection schemes
of wide usage in computational fluid dynamics and numerical astrophysics,
{\it e.g.} the Flux Corrected Transport algorithm of \inlinecite{Boris1973}
(and its many subsequent variants and elaborations),
and the PPM scheme of \inlinecite{Woodward1984}.
In such schemes, dissipation
is introduced at the algorithmic level, rather than as explicit
dissipation terms in the governing equations, as in conventional Large
Eddy Simulation
approaches. The crux in the design of ILES scheme is to keep this
numerical dissipation at a minimum, {\it i.e.} to ensure 
it activates only when and where
it is needed to maintain locally smooth ({\it i.e.} spurious oscillation free) solution.
In such schemes the level of implicit
diffusivities typically decreases with increasing mesh size, but the
absolute level of numerical dissipation also depends on the specific
algorithm used. 
Numerous examples are provided in the volume edited
by \inlinecite{Grinstein2007}, including applications to local and global
solar--stellar convection.

One drawback of ILES schemes is the
difficulty in calculating, even {\it a posteriori}, conventional dimensionless
numbers such as the Reynolds or Rayleigh numbers, which complicates
comparison with simulations carried out including explicit dissipation.

From the practical point of view, however, their main advantage 
is that nonlinearly stable turbulent simulations can be produced on relatively coarse
grids (\opencite{Smolarkiewicz2002}; \opencite{Domaradzki2003};
\opencite{Margolin2006b}; \opencite{Smolarkiewicz2007}),
which, in particular, allows long temporal integration, as required in the study 
of behaviors such as magnetic cycles, developping on timescales very much longer
than the turbulent turnover time.

The specific simulation segment analyzed in what follows spans 180 years, and is executed
on a relatively small mesh of size
$N_r\times N_\theta\times N_\phi=47\times 64\times 128$ in
radius $\times$ latitude $\times$ longitude. The spatial domain
is a thermally forced thick spherical shell ($0.602\leq r/R\leq 0.96$) rotating initially
rigidly at the solar rate, convectively unstable in its outer two thirds
($0.718\leq r/R\leq 0.96$). Small perturbations in radial flow speed and
magnetic field are introduced in the convectively unstable portion
of the domain at $t=0$.

The foregoing analysis begins with the four-dimensional datasets (three spatial
dimensions plus time) returned by the simulation. The first step is to
extract the axisymmetric components of the total flow and magnetic field.
As shown by \inlinecite{Racine2011} through modal decomposition,
these axisymmetric components evolve
on a timescale much longer than their non-axisymmetric counterparts,
and can thus be legitimately considered as a distinct dynamical entity.
We therefore express the total flow [${\bf U}$] and magnetic field [${\bf B}$]
as
\begin{eqnarray}
\label{scU}
{\bf U}(r,\theta,\phi,t) &=& {\bf u} (r,\theta,t) + {\bf u'}(r,\theta,\phi,t) ~, \\
\label{scB}
{\bf B}(r,\theta,\phi,t) &=& {\bf b} (r,\theta,t) + {\bf b'}(r,\theta,\phi,t) ~,
\end{eqnarray}
where
\begin{eqnarray}
{\bf u}(r,\theta,t) &=& \langle {\bf U} (r,\theta,\phi,t) \rangle ~, \\
{\bf b}(r,\theta,t) &=& \langle {\bf B} (r,\theta,\phi,t) \rangle ~,
\end{eqnarray}
are the axisymmetric large-scale components, calculated by zonal averaging:
\begin{equation}
\label{avg}
\langle{\bf X} (r,\theta,\phi,t) \rangle
= \frac{1}{2\pi}\int_0^{2\pi} {\bf X}(r,\theta,\phi,t)\,{\rm d}\phi~.
\end{equation}
Note that under these definitions,
$\langle {\bf u'} \rangle = \langle {\bf b'} \rangle = 0$,
so that the non-axisymmetric contributions of the flow and field play the role
of the ``small scales'' in mean-field theory.

\begin{figure}[p!]
\begin{center}
\includegraphics[width=1.0\linewidth]{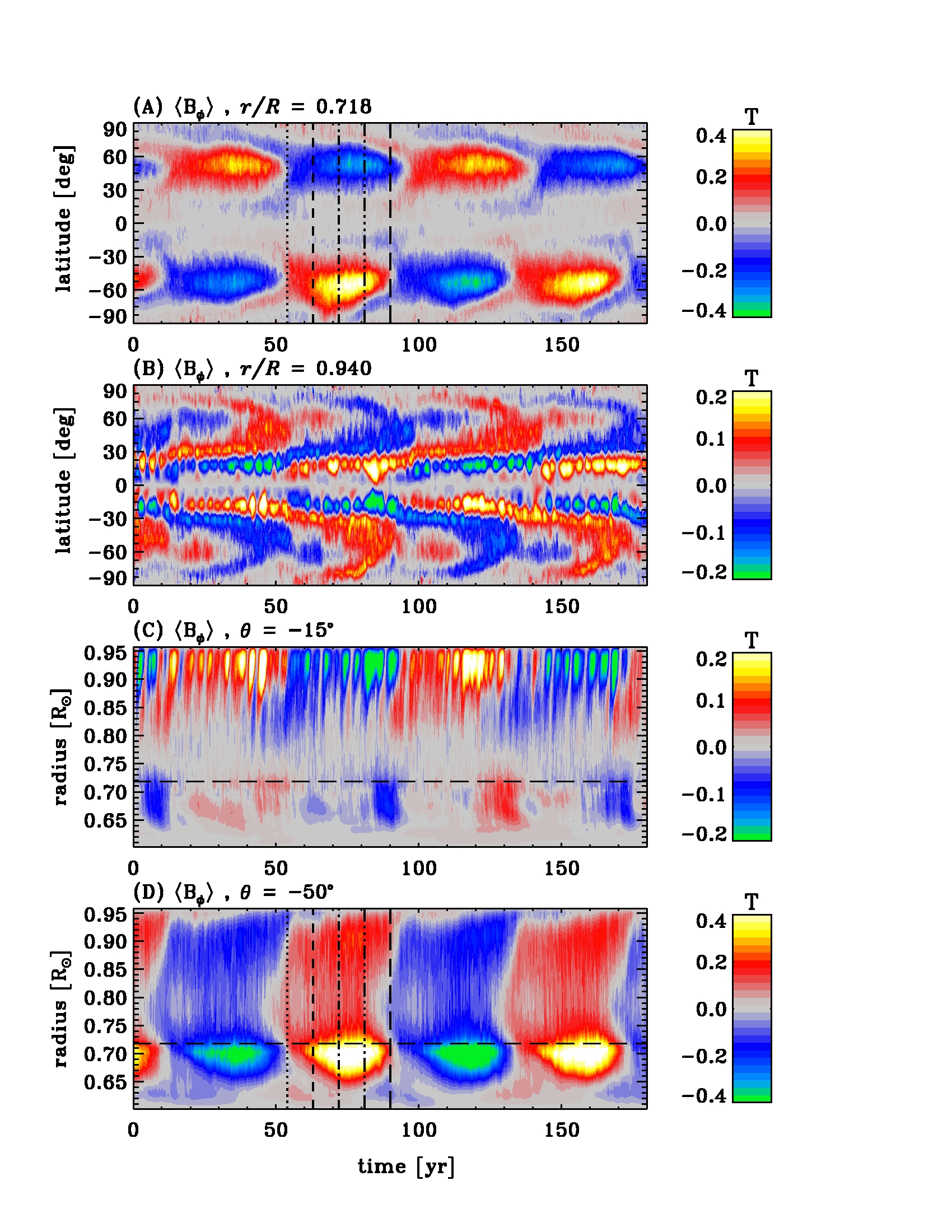}
\caption{Two latitude--time representations of the zonally averaged
toroidal magnetic field
at two different radii, along with two radius--time representations at
two different latitudes. The dashed line in panels (c) and (d) represents the interface
between the convectively stable and unstable fluid layers in the simulation's
background stratification. The vertical lines in panels (a) and (d) flag five
specific epochs across the second half-cycle, for subsequent reference.
}\label{tor_mag}
\end{center}
\end{figure}

Figure~\ref{tor_mag} offers four views of the large-scale (axisymmetric)
toroidal magnetic-field component evolving over the timespan of the simulation. The
top two panels (a) and (b) show time--latitude cuts, the first extracted at the depth
coinciding with the base of the convecting layers
($r/R=0.718$), and the other near its top ($r/R=0.94$). The bottom two panels
show time--radius cuts extracted at (c) low and (d) mid-latitudes in the southern
hemisphere. Regular polarity reversals of the large-scale magnetic field
stand out prominently in these diagrams, here on a period of about 36 years
for each half cycle (equivalent to a sunspot cycle); thus the magnetic
cycle period in this simulation is a little over three times longer
than the $\approx 22\,$years observed on the Sun. The large-scale toroidal
component is antisymmetric about the equatorial plane, in agreement with
Hale's polarity laws, and peaks at mid-latitudes (panel (a)) and immediately
beneath the core--envelope interface (panel (d)); this latter property is
in line with
the need to form and store in the convectively stable
layer the toroidal magnetic flux
ropes that, upon buoyancy-driven destabilization and emergence,
will give rise to sunspots
(see \opencite{Fan2009}, and references therein). The subsurface time--latitude
diagram on panel (b) and time--radius diagram on panel (c)
also show a hint of a secondary dynamo mode,
of much shorter period and lower amplitude than the primary mode,
producing what looks like an oscillation
superimposed on the more slowly evolving magnetic component pervading
the bulk of the domain. Interestingly, a similar combination of
long- and short-period dynamo modes was also observed in the spherical wedge
simulations of \inlinecite{Kapyla2010}. This intriguing dynamo
feature will be revisited in what follows.

Other features of this simulation are discrepant
with respect to the solar cycle, besides the period.
Most notably, the toroidal magnetic component
at the core--envelope interface, where sunspots are presumed to originate,
peaks at too high latitudes compared to the sunspot butterfly diagram,
and only shows a hint of equatorward migration.
Moreover, analysis of the poloidal large-scale component reveals that the
latter oscillates in phase with the deep-seated toroidal component, whereas
in the Sun a phase lag of $\pi/2$ is inferred. The specific simulation we are
using for the foregoing analysis develops a slow phase drift between hemispheres,
which eventually leads, after some 300 years, to a switch to a non-axisymmetric
large-scale dynamo mode, a fascinating dynamo behavior in and of itself.
Despite these departures with respect to observed solar behavior,
the presence
of a well-defined cyclic behavior in the large-scale magnetic field offers a
unique opportunity
to investigate quantitatively the magnetic back reaction on
large-scale flows building up in the simulation and observed in the Sun,
in particular differential rotation.

\subsection{The Mean Differential Rotation}

Mechanically speaking,
solar differential rotation is driven primarily by Reynolds stresses arising through
rotation-driven anisotropies in convective turbulence and angular-momentum transport
by meridional flows. Helioseismology has now mapped, with good accuracy,
differential rotation throughout the bulk of the solar convection zone and upper
radiative core ({\it e.g.} Figure~18 in \opencite{Howe2009}). If one excludes
subphotospheric layers, the primary rotational gradient in the solar convection
zone is latitudinal, with the rotational frequency of equatorial regions
exceeding that of polar regions by about 30\%. This latitudinal gradient
vanishes at the interface between the convection zone and the underlying radiative
core, across a thin shear layer known as the tachocline.

\begin{figure}[p!]
\begin{center}
\includegraphics[width=1.0\linewidth]{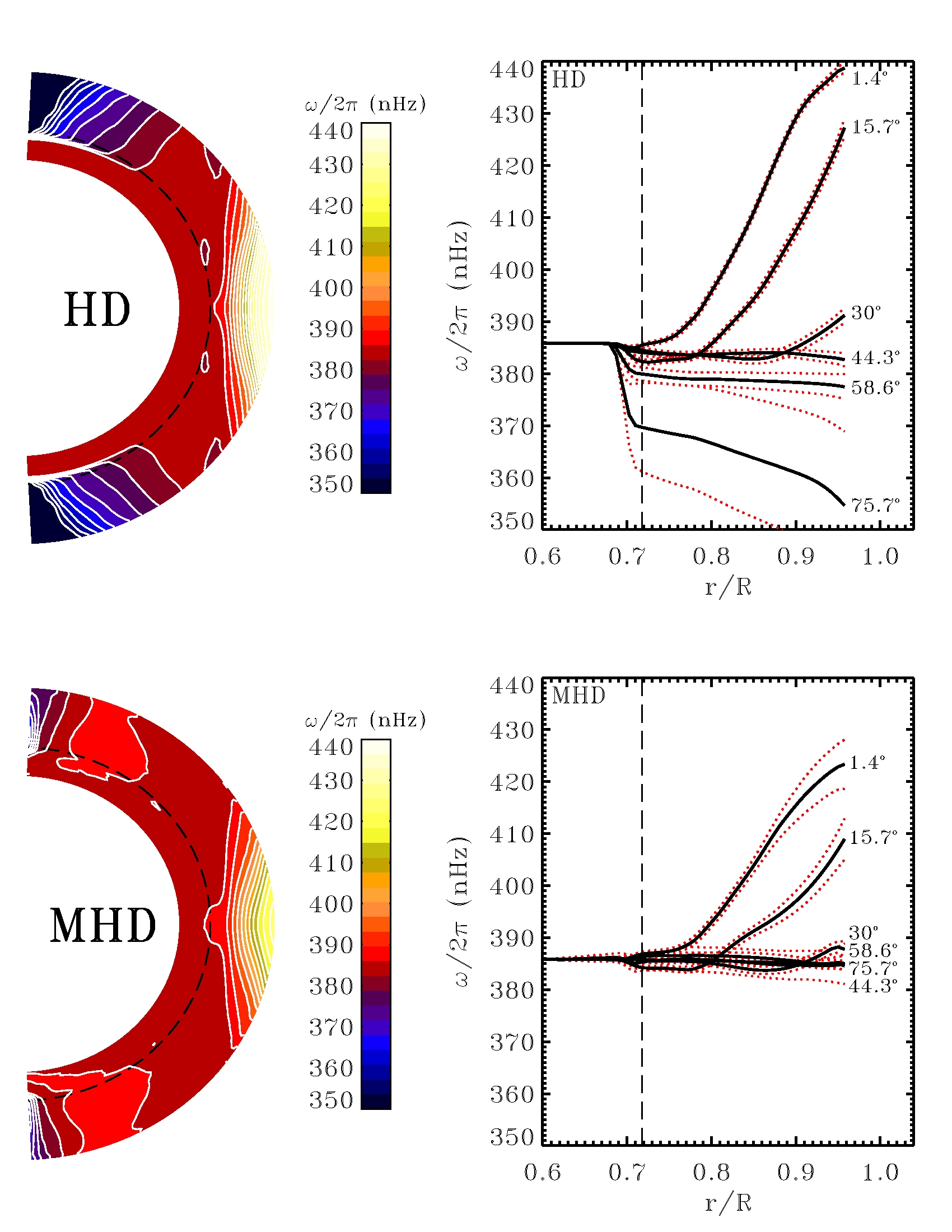}
\caption{Two different representations of the angular velocity in a purely hydrodynamical
simulation (top)
and the MHD simulation of Figure~\ref{tor_mag} (bottom). The left panels show
contour plots of rotational
frequency at each position in the Sun, constructed from the zonally averaged
longitudinal velocity averaged in time over the duration of the simulation.
The right panels show the same results in the form of constant-latitude
radial cuts in the northern hemisphere,
highlighting the presence of a tachocline-like shear layer
immediately beneath the core--envelope interface ($r/R=0.718$, dashed line).
The red dotted curves indicate the $\pm\,1$ root-mean-square deviations about
the temporal mean.}\label{difrot}
\end{center}
\end{figure}

With simulations computed in a reference frame rotating at the mean solar
rate ($2.42405 \times 10^{-6}~\mathrm{rad}\,\mathrm{s}^{-1}$), differential rotation 
can be directly computed from
the zonally averaged $\phi$-component of the flow velocity. We first
compute the mean differential rotation profile by temporally averaging
over the full temporal extent of the simulation, which is in a
statistically equilibrated state.
We have carried out this averaging exercise for the MHD simulation of
Figure~\ref{tor_mag}, as well as a purely HD convection simulation
operating under the same forcing regime and rotation rate, and computed using
the same mesh size.
The results are shown on Figure~\ref{difrot}, expressed as angular velocity
$\omega=\langle u_\phi\rangle/(r\cos\theta)$, with $-\pi/2\leq \theta\leq \pi/2$ the
latitude.
The left panels show isocontour maps, with corresponding radial cuts
plotted on the right panels, on the same scales 
to allow quick visual comparison of the two simulations.
Both are characterized by equatorial acceleration, but with
isocontours too closely aligned with the rotation axis, and too concentrated
towards the middle of the convection zone at low latitudes, as compared
to the helioseismically-inferred solar internal differential rotation.
These features are
in fact typical of these types of simulations ({\it e.g.} Figure~9 in \opencite{Brun2004};
Figure~1 in \opencite{Browning2006};
Figure~3 in \opencite{Brown2008}), unless a latitudinal gradient in the heat flux is artificially
imposed at the base of the domain (see \opencite{Miesch2006}).
Nonetheless,
the differential rotation characterizing the
HD simulation (top row) shows some remarkably solar-like features, notably the
magnitude of
the Pole-to-Equator angular velocity contrast,
and, in particular, a thin tachocline-like rotational shear layer
located immediately beneath the core--envelope interface. The thinness
of this layer ($\approx 0.025 R$ here) is a direct reflection of the
very low dissipation levels characterizing this simulation ({\it cf.}~Figure~1
in \opencite{Browning2006}).

The resemblance to solar differential rotation degrades, however, upon
moving to the MHD simulation (bottom row in Figure~\ref{difrot}). The Pole-to-Equator
angular velocity contrast is now reduced by a factor of three as compared to
the HD simulation, and the latitudinal gradient has all but vanished
at mid to high latitudes. Such strong magnetic backreaction on the mean differential
rotation is in fact typical of these types of global MHD convection simulations,
as shown already by \inlinecite{Gilman1981} and \inlinecite{Gilman1983}. 
A residual tachocline remains, in the sense that
the weak convection zone latitudinal differential rotation again vanishes
across a thin shear layer beneath the core--envelope interface.
Although quite weak, this differential rotation remains important
for the operation of the dynamo, as the analysis of a similar
simulation carried out in \inlinecite{Racine2011} shows that it contributes
approximately equally with the turbulent electromotive force to the
regeneration of the large-scale toroidal magnetic field near the
core--envelope interface. 

Another important difference between the differential rotation profiles
characterizing the HD and MHD simulations is that the former is temporally
steady once the simulation has attained a statistically stationary state.
The dotted lines bracketing each radial cut on the right panels of Figure~\ref{difrot}
indicate the $\pm 1 \sigma$ deviations of the zonal averages about their temporal
mean plotted on the left panels. These are quite small in the HD simulation,
except near the poles.
The MHD simulation, on the other hand, exhibits somewhat larger one-sigma
deviations at most latitudes, but these now reflect the presence of
spatiotemporally coherent cyclic variations
superimposing themselves on the mean rotational profile. We now turn to the
characterization of these torsional oscillations.

\subsection{The Torsional Oscillations}

\begin{figure}[h!]
\begin{center}
\includegraphics[width=0.95\linewidth]{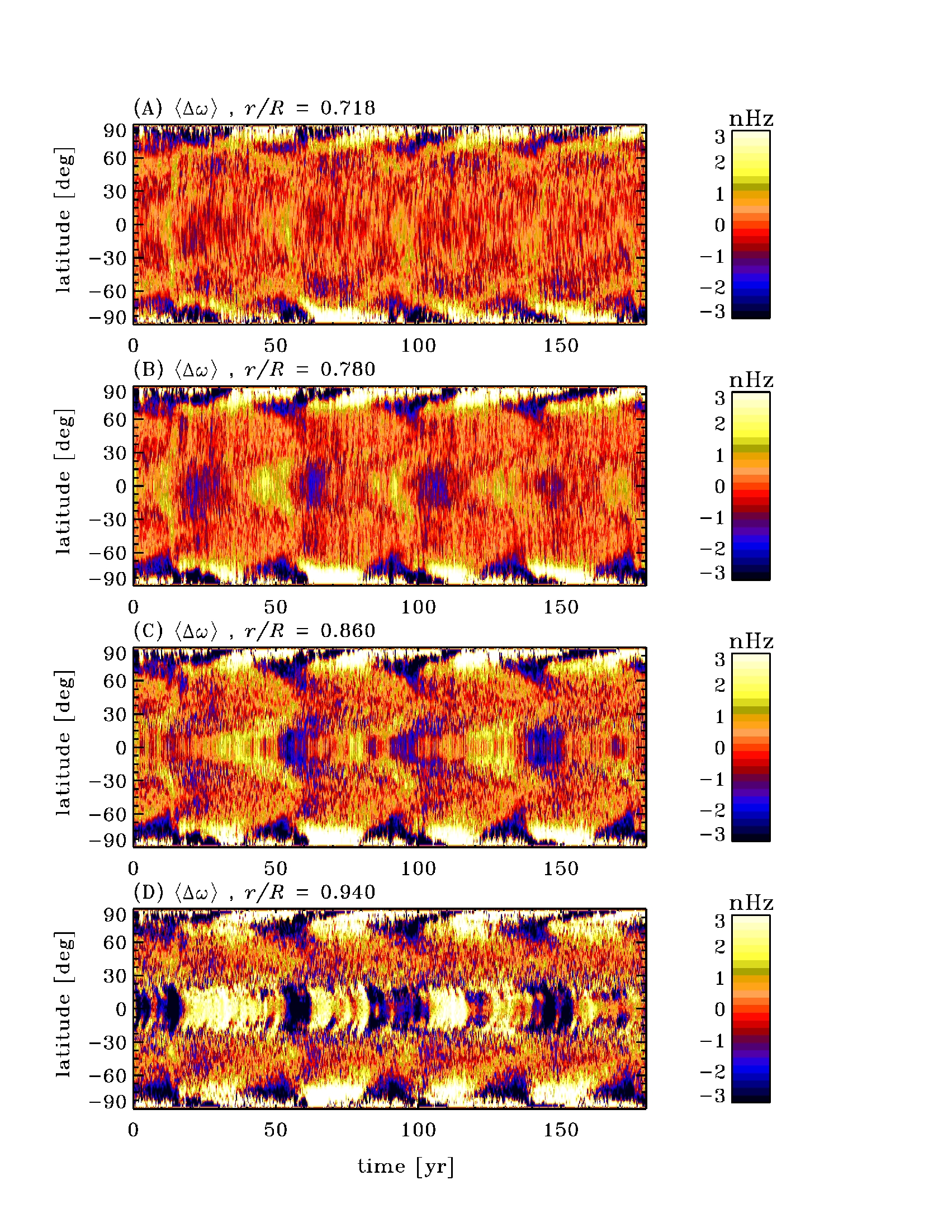}
\caption{Time--latitude diagrams of the zonally averaged
perturbation in rotational frequency, at four different depths in the simulation.
Positive (negative) perturbations correspond to
rotational acceleration (deceleration) with respect to the mean state plotted
on Figure~\ref{difrot}(b)}\label{u_t}
\end{center}
\end{figure}

\begin{figure}[h!]
\begin{center}
\includegraphics[width=0.95\linewidth]{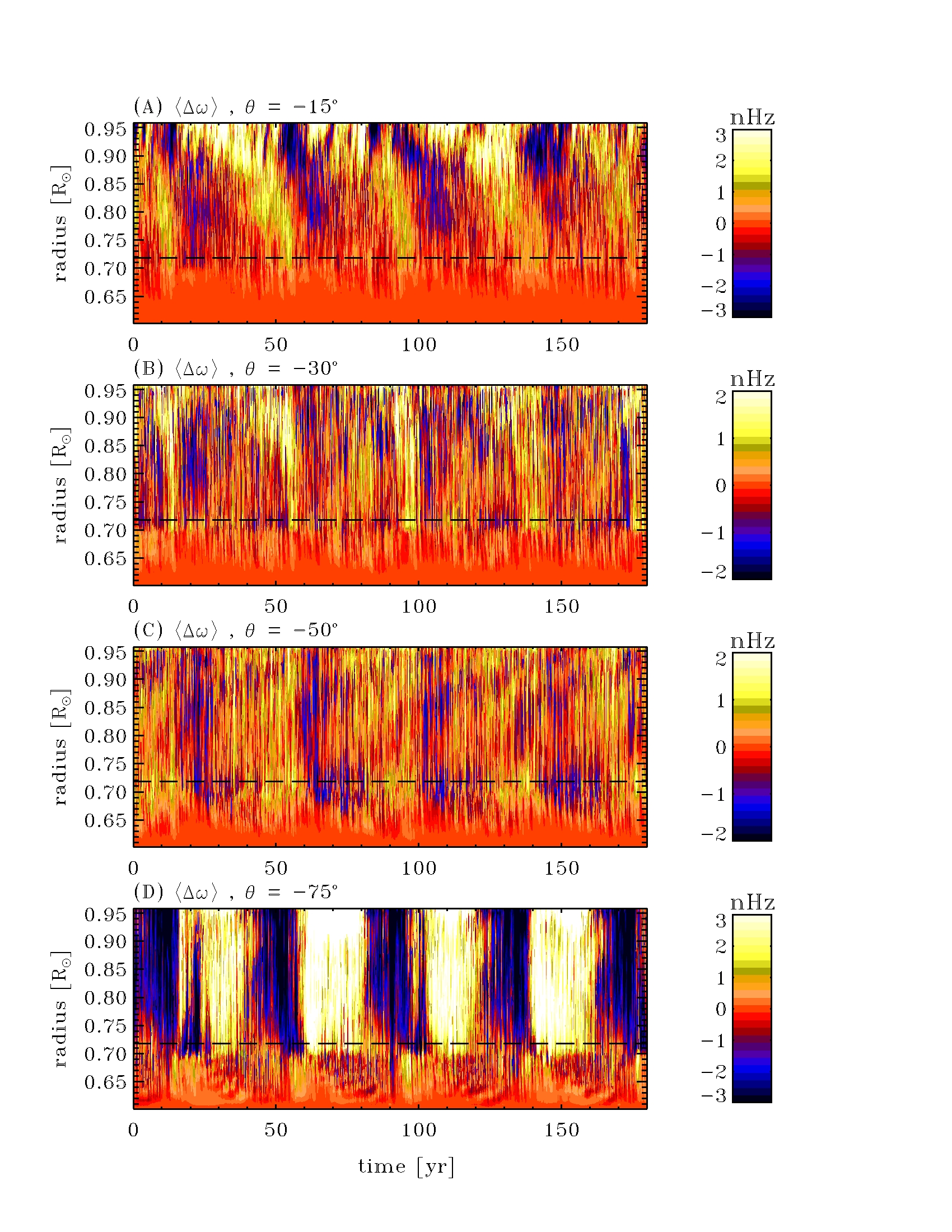}
\caption{Time--radius diagrams of the zonally averaged
perturbation in rotational frequency, at four different latitudes in the simulation.
The dashed line indicates the position
of the core--envelope interface.}\label{u_r}
\end{center}
\end{figure}

Torsional oscillations are best visualized by subtracting the temporally averaged
rotational frequency profile of Figure~\ref{difrot} from the corresponding
zonally-averaged rotational frequency at each time step:

\begin{equation}
\langle\Delta\omega\rangle(r,\theta,t)= (2\pi r\cos\theta)^{-1}
(u_{\phi} (r,\theta,t) - \bar{u}_{\phi} (r,\theta))~,
\end{equation}
where the overbar denotes temporal averaging over the timespan of the simulation:
\begin{equation}
\bar{u}_{\phi} (r,\theta) = \frac{1}{2 \pi T} \int_0^{T} \int_0^{2 \pi} U_{\phi}(r,\theta,\phi,t) \,\mathrm{d} \phi \, \mathrm{d} t ~,
\end{equation}
amounting to the zonal and temporal average of the full azimuthal velocity
[$U_{\phi}(r,\theta,\phi,t)$] numerical data set, with $T$ the length
of the simulation.
The result of this procedure is shown on Figures
\ref{u_t} and \ref{u_r}, which display respectively time--latitude diagrams at four fixed
depths, and radius--time diagrams at four fixed latitudes.
Several features visible in these plots are noteworthy:
i) A cyclic signal is clearly present at all depths and latitudes so sampled,
at twice the frequency characterizing the magnetic cycle (cf.~Figure~\ref{tor_mag});
ii) The torsional oscillations peak in amplitude at high latitudes and in the surface
and subsurface layers, reaching there $\approx\,3\,$nHz; iii)
At mid- to high latitudes, the oscillations show a phase approximately independent of depth.
The oscillations reach their peak prograde phase ({\it i.e.} $\langle\Delta\omega\rangle$
peaking at positive values) at about the peak of the magnetic cycle.
All of these features are remarkably solar-like, as can be inferred from comparison
with a similarly formatted diagram in Howe (2009; {\it cf.}~Figure~26 to Figure~\ref{u_r}
herein).

At first glance, the surface spatiotemporal pattern of the torsional oscillations
is not particularly solar-like ({\it cf.}~Figure~25 of \opencite{Howe2009}, and Figure~\ref{u_t}(d) herein).
Near the surface, a strong and rather complex oscillatory signal is present at
low latitudes, arising from the superimposition of an oscillation associated with
the large-scale magnetic cycle with a second, characterized by higher frequency
and restricted to the subsurface equatorial regions. Examination of the simulation
reveals that this is associated with a secondary dynamo mode feeding on
the strong latitudinal shear present in the outer half of the convection zone
at low latitudes (see Figure~\ref{difrot}; also Figure~\ref{tor_mag}(b) and (c)).

\begin{figure}[h!]
\begin{center}
\includegraphics[width=0.95\linewidth]{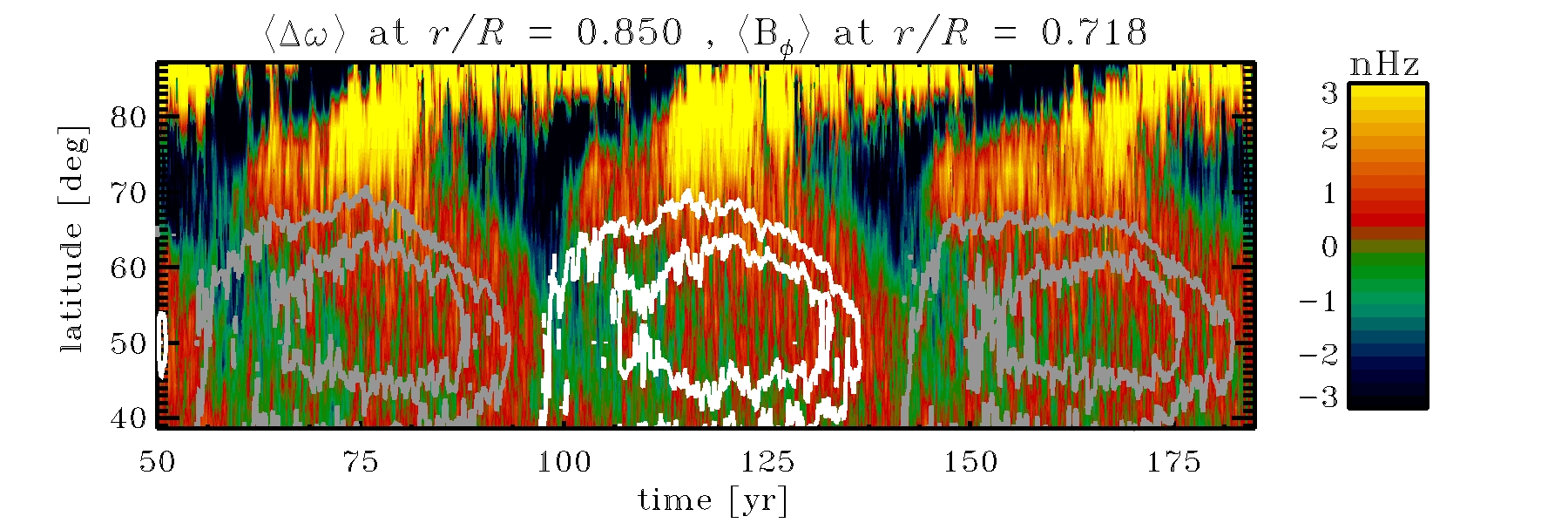}
\caption{A close up on the time--latitude diagram of Figure~\ref{u_t}(d), showing the
zonally averaged rotational frequency perturbation. The color scale and range
are now chosen identical to those used in Figure~25 of Howe (2009), to ease
comparison. A few isocontours of zonally averaged toroidal field at the core--envelope
interface ($r/R=0.718$) have been superimposed, to indicate the relative phase of
the magnetic cycle. White (grey) correspond to positive (negative) toroidal field,
and the innermost contours have values $\pm$~0.2~T.}\label{tors_bx}
\end{center}
\end{figure}

However, the dissimilarities greatly diminish if one focuses on the pattern
present at latitudes higher than the equatorial magnetic ``activity belts'',
as defined by the time--latitude distribution of the large-scale zonal magnetic component
at the core--envelope interface (Figure~\ref{tor_mag}(a)), where sunspots are presumed
to originate.
Figure \ref{tors_bx} illustrates the idea. It is essentially a closeup of the
northern hemisphere portion of Figure~\ref{u_t}(c), on which have been superimposed
a few isocontours of mean toroidal magnetic field strengths at the core--envelope
interface ({\it cf.}~Figure~\ref{tor_mag}(a)) for the first three half-cycles in the simulation.
This stretching procedure ``renormalizes'' the activity belts to low
latitudes, and allows a comparison that is arguably more relevant to the patterns observed
on the Sun. If one accepts this stretch at face value, then the comparison becomes actually
quite good ({\it cf. e.g.}~Figure~25 in \opencite{Howe2009}). In particular, the torsional
acceleration ($\langle\Delta\omega\rangle>0$)
is seen to begin at high latitudes
(here $\approx\pm~70^\circ$) at about the time of magnetic-polarity
reversal (akin to solar minimum here), and develops in two diverging branches, one
propagating poleward and the second, of lower amplitude, propagating equatorward.
In the Sun, this second branch requires two activity cycles to reach the
Equator, while here it does so in only one cycle; this may be a reflection
of the fact that our activity belts are located at too high latitudes, but
this remains to be demonstrated through further simulations.
It is also quite remarkable that this latitudinal double-branch pattern
in the torsional oscillations persists all the way to the base
of the convecting layers in the simulation. Careful examination of
Figure~\ref{u_r}(d) also reveals that an oscillatory signal even penetrates
all the way through the underlying stably stratified layer to the base
of the computational domain ($r/R=0.602$), although
this signal may reflect the excitation of gravity waves.

At any rate, the results presented here demonstrate clearly that
this simulation generates a global torsional oscillation pattern
that is solar-like in a number of ways. Remaining discrepancies
notwithstanding, we have in hand a unique ``virtual laboratory'' allowing
a quantitative and fully dynamical investigation of the mechanism(s)
driving these torsional ocillations. This is the topic to which we now turn.

\section{The Dynamical Drivers of Torsional Oscillations}

\subsection{The Zonal Momentum Equation}

Torsional oscillations are generated because the (magneto)dynamics of the convecting
layers lead to a systematic, time-dependent redistribution of angular momentum that
is driven, directly or indirectly, by the magnetic cycle. In parallel to the
MHD equations solved in the simulation, the foregoing analysis
focuses on the azimuthal component of the Navier--Stokes equations,
including the Lorentz and Coriolis forces, and written under the anelastic
approximation. The latter implies, in particular, that mass conservation
reduces to $\nabla \cdot (\rho {\bf U}) = 0$, with $\rho = \rho(r)$.

The starting point of our analysis is to recast the azimuthal momentum
equation in conservative
form involving fluxes of angular momentum.
We begin by applying
the azimuthal averaging operator previously defined in Equation~(\ref{avg}):
\begin{equation}
\left \langle \frac{\partial (\rho U_{\phi})}{\partial t} \right \rangle + \langle [\nabla \cdot (\rho {\bf U} {\bf U})]_{\phi} \rangle = -\frac{1}{r \cos \theta} \left \langle \frac{\partial P}{\partial \phi} \right \rangle + \frac{1}{\mu_0} \langle [(\nabla \times {\bf B}) \times {\bf B}]_{\phi} \rangle~,
\end{equation}
with $P$ the pressure profile, ${\bf U} = {\bf V} + {\bf \Omega} \times {\bf R}$,
where ${\bf V}$ is the velocity
field of the plasma in the rotating frame of the Sun, ${\bf \Omega}$ is the mean angular velocity
of the Sun, ${\bf R}$ is a radial vector locating a given fluid element in a spherical
coordinate system with origin at the Sun's center, and once again the variable $\theta$
denotes the latitude. Now, the azimuthal averaging operator commutes with derivatives
acting on large-scale quantities, and the Lorentz
force term can be rewritten as
$(\nabla \times {\bf B}) \times {\bf B}
= ({\bf B} \cdot \nabla) {\bf B} - \frac{1}{2} \nabla {\bf B}^2$.
Moreover, under the axisymmetry assumption,
all azimuthal derivatives of averaged quantities automatically vanish, so that
the above expression reduces to
\begin{equation}
\label{red}
\frac{\partial \langle \rho U_{\phi} \rangle}{\partial t} + \langle \hat{\bf e}_{\phi} \cdot [\nabla \cdot (\rho {\bf U} {\bf U})] \rangle = \frac{1}{\mu_0} \langle \hat{\bf e}_{\phi} \cdot ({\bf B} \cdot \nabla) {\bf B} \rangle~.
\end{equation}
Use of judicious vector identities also leads to
\begin{eqnarray}
\label{vU}
\langle \hat{\bf e}_{\phi} \cdot [\nabla \cdot (\rho {\bf U} {\bf U})] \rangle &=& \frac{1}{r \cos \theta} \nabla \cdot (r \cos \theta \rho \langle U_{\phi} {\bf U} \rangle) ~, \\
\label{vB}
\langle \hat{\bf e}_{\phi} \cdot [({\bf B} \cdot \nabla) {\bf B}] \rangle &=& \frac{1}{r \cos \theta} \nabla \cdot (r \cos \theta \langle B_{\phi} {\bf B} \rangle) ~.
\end{eqnarray}

Using the scale separation introduced in Equation~(\ref{scU}) -- (\ref{scB}), the terms
$\langle U_{\phi} {\bf U} \rangle$ and $\langle B_{\phi} {\bf B} \rangle$
can be separated between large-scale and small-scale contributions, {\it e.g.}:
\begin{equation}
\langle U_{\phi}{\bf U} \rangle = \langle (u_{\phi} + u'_\phi)({\bf u}+{\bf u'}) \rangle 
= u_{\phi} {\bf u} + \langle u'_{\phi} {\bf u'} \rangle~,
\end{equation}
and likewise for $\langle B_{\phi}{\bf B} \rangle$. Inserting Equation~(\ref{vU}) and Equation~(\ref{vB}) into Equation~(\ref{red}) yields
\begin{equation}
\label{almost}
\frac{\partial (\rho u_{\phi})}{\partial t} - \frac{1}{r \cos \theta} \nabla \cdot \Big\{ r \cos \theta \Big[ \frac{1}{\mu_0} ( b_{\phi} {\bf b} + \langle b'_{\phi} {\bf b'} \rangle) - \rho ( u_{\phi} {\bf u} + \langle u'_{\phi}{\bf u'} \rangle) \Big] \Big\} = 0~.
\end{equation}
One can separate the rigid rotation component by writing 
${\bf u} = {\bf v} + {\bf \Omega} \times {\bf r}$, noting that ${\bf u'} = {\bf v'}$
because there is no small-scale contribution to ${\bf \Omega} \times {\bf R}$. Since the latter
term is also divergence free, we can write
\begin{equation}
u_{\phi} {\bf u} = (v_{\phi} + \Omega r \cos \theta) {\bf v}~.
\end{equation}
The end result of all this is to recast Equation~(\ref{almost}) in conservative form: 
\begin{eqnarray}
\label{eq:cons}
\frac{\partial (\rho v_{\phi})}{\partial t} &-& \frac{1}{r \cos \theta} \nabla \cdot \Big\{ r \cos \theta \Big[ \frac{1}{\mu_0} ( b_{\phi} {\bf b} + \langle b'_{\phi} {\bf b'} \rangle) \nonumber \\
& & -\, \rho (( v_{\phi} + \Omega r \cos \theta) {\bf v} + \langle v'_{\phi}{\bf v'} \rangle) \Big] \Big\} = 0~.
\end{eqnarray}
The four terms adding up under the divergence on the LHS act as volumetric force densities
in the longitudinal direction, and define here
the four contributors to zonal dynamics:
\begin{eqnarray}
\label{FReyn}
F_{\mathrm{Reyn}} &=& \frac{-1}{r \cos \theta} \nabla \cdot (r \cos \theta \,\rho \langle v'_{\phi}{\bf v'} \rangle) ~, \\
\label{FCirc}
F_{\mathrm{Circ}} &=& \frac{-1}{r \cos \theta} \nabla \cdot (r \cos \theta \,\rho( v_{\phi} + \Omega r \cos \theta) {\bf v}) ~, \\
\label{FMaxw}
F_{\mathrm{Maxw}} &=& \frac{1}{r \cos \theta} \nabla \cdot \Big(\frac{r \cos \theta}{\mu_0} \langle b'_{\phi} {\bf b'} \rangle \Big) ~, \\
\label{FMagn}
F_{\mathrm{Magn}} &=& \frac{1}{r \cos \theta} \nabla \cdot \Big(\frac{r \cos \theta}{\mu_0} \, b_{\phi} {\bf b} \Big) ~.
\end{eqnarray}
These are, respectively,
turbulent Reynolds stresses, Coriolis force acting on the meridional flow, turbulent Maxwell
stresses (small-scale Lorentz force) and the magnetic torque (Lorentz force associated with
the large-scale magnetic component).
Note that viscous stresses do not appear explicitly here, as our simulation
is of the implicit large-eddy type, {\it i.e.} it does not include explicit dissipative
terms in the momentum equation. 

\subsection{Angular-Momentum Fluxes}

As per Equation~(\ref{eq:cons}), the mean radial and latitudinal angular-momentum fluxes
are given by
\begin{eqnarray}
\mathcal{L}_r(r,\theta,t) &=& r \cos \theta \Big( \frac{1}{\mu_0} ( b_{\phi} b_r + \langle b'_{\phi} b'_r \rangle) \nonumber \\
& & -\, \rho (( v_{\phi} + \Omega r \cos \theta) v_r + \langle v'_{\phi} v'_r \rangle) \Big) ~, \\
\mathcal{L}_{\theta}(r,\theta,t) &=& -r \cos \theta \Big( \frac{1}{\mu_0} ( b_{\phi} b_{\theta} + \langle b'_{\phi} b'_{\theta} \rangle) \nonumber \\
& & -\, \rho (( v_{\phi} + \Omega r \cos \theta) v_{\theta} + \langle v'_{\phi} v'_{\theta} \rangle) \Big) ~, 
\label{Ltheta}
\end{eqnarray}

Following \inlinecite{Brun2004}, we first examine the global rotational dynamics
by computing from the simulation output the fluxes
of angular momentum integrated across spherical shells or conical wedges
centered on the rotation axis:
\begin{eqnarray}
I_r(r,t) &=& \int_{-\pi/2}^{\pi/2} \mathcal{L}_r(r,\theta,t) r^2 \cos \theta \,\mathrm{d}\theta ~,
\label{Ir}
 \\ 
I_{\theta}(\theta,t) &=& \int_{r_{bot}}^{r_{top}} \mathcal{L}_{\theta}(r,\theta,t) r \cos \theta \,\mathrm{d}r ~,
\label{Itheta}
\end{eqnarray}
so that $I_r(r,t)$ is the net angular-momentum transport rate through shells of different radii,
and $I_{\theta}(r,t)$ through cones tangent to different latitudes.
 
\begin{figure}[p!]
\begin{center}
\includegraphics[width=0.95\linewidth]{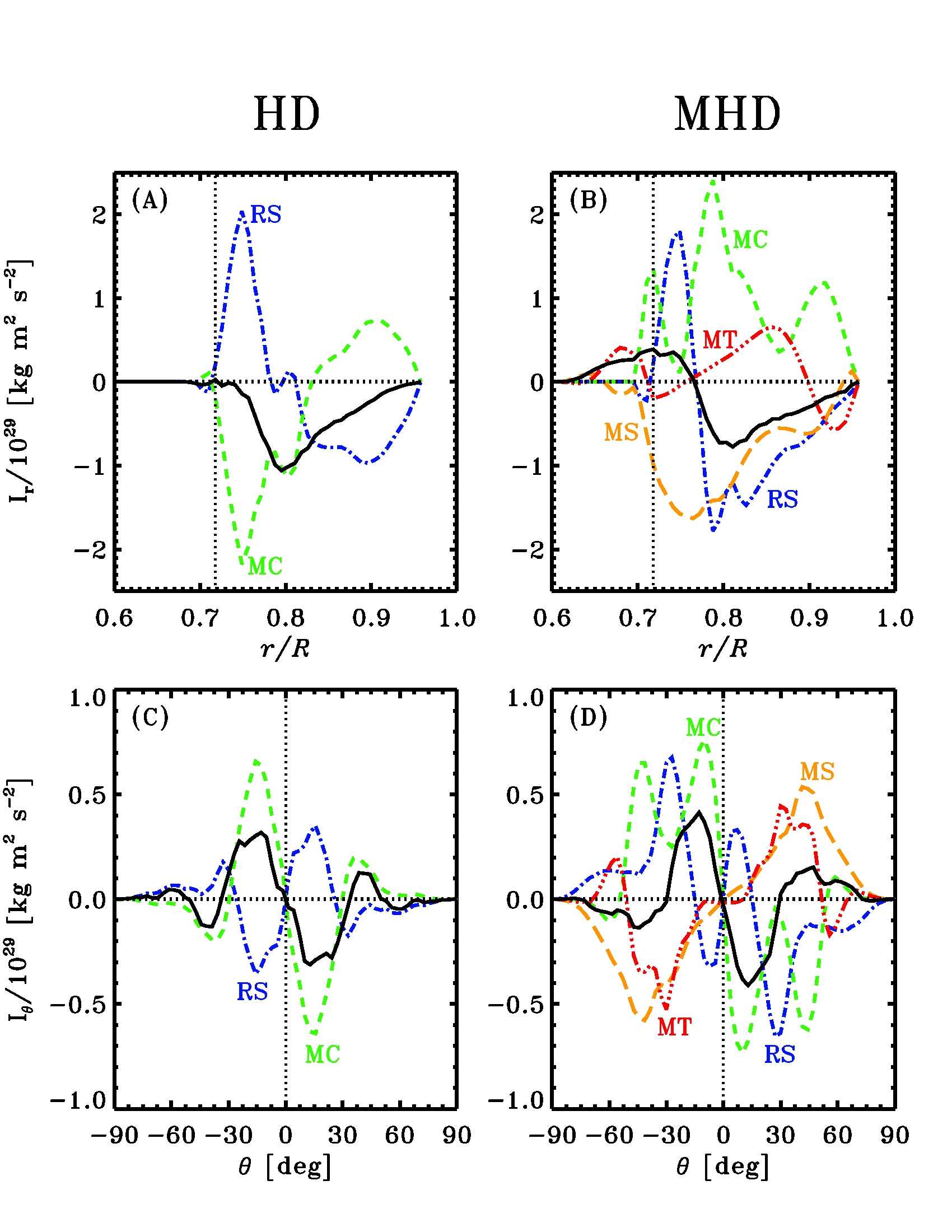}
\caption{Angular-momentum flux contributions.
Panels (a) and (c) show these contributions in the HD simulation, while panels (b) and (d)
show them in the MHD simulation. Panels (a) and (b) represent angular-momentum fluxes
through spherical shells of different radii, with positive values amounting to
upward transport of angular momentum. Panels (c) and (d) show the fluxes
through conical surfaces of different latitudes, with positive values indicating
northward transport. The various curves are color-coded (line-styled)
as follows: blue (dot-dashed) -- Reynolds stresses; green (short-dashed) -- Coriolis force; orange (long-dashed) -- Maxwell stresses;
red (triple-dot-dashed) -- large-scale magnetic fields; black (full) -- total of all the contributions.
The vertical dotted lines in the top two panels show the location of the core--envelope
interface at $r/R = 0.718$.}\label{flux_plots}
\end{center}
\end{figure}

In order to disentangle the various physical contributions to angular-momentum
transport, these integrals are computed separately for the four distinct contributions
to the total angular-momentum fluxes. The result of this procedure is shown in
Figure~\ref{flux_plots}, where the fluxes have also been temporally averaged
over the extent of the simulation. The procedure was also carried out for the same
parent HD simulation whose mean rotational frequency is plotted on
Figure~\ref{difrot} (top). In this HD simulation, the only contributors to zonal
dynamics are the Reynolds stresses and Coriolis force acting on the meridional flow.

In the HD simulation, the Reynolds stresses and Coriolis force are seen to act
in opposition at all latitudes and depths. This is precisely what one would expect
for a stationary rotational state to ensue. With all fluxes vanishing
at domain boundaries, the fact that these two contributions
do not add up to zero reflects the presence, in the simulation, of a dissipative
force associated with the numerical scheme, which here contributes
to the zonal dynamics, especially at low latitudes in the convection zone.
This
can be traced in part to the very strong rotational shear which builds up there 
in the HD simulation
({\it cf.}~top panels in Figure~\ref{difrot})\footnote{Such a dissipative force,
of purely numerical origin, is the hallmark of ILES simulations.
In the specific context of EULAG, the hydrodynamical parent of EULAG--MHD,
it has been shown by \inlinecite{Margolin1999}
to mimic {\it bona fide} subgrid scale parametrization in the context
of the planetary boundary layer,
and, more importantly, to effectively vanish
if an appropriate explicit subgrid model is introduced in the model.}

Turning to the MHD simulation, the most obvious feature is perhaps the fact
that all four potential fluxes of angular momentum contribute more
or less equally to the global rotational dynamics. Moreover, the presence of
magnetic fields has greatly altered the non-magnetic fluxes of angular momentum.
This is particularly the case for the radial flux of angular momentum associated
with the Coriolis force ({\it cf.}~green curves on panels (a) and (b)), and latitudinal
fluxes by Reynolds stresses
({\it cf.}~blue curves on panels (c) and (d)), which both undergo reversals in direction
over a substantial portion of their spatial range when going from HD to MHD.
Also noteworthy in this MHD simulation, the Maxwell stresses associated with
the small-scale magnetic field contribute more or less equally to the Lorentz
force associated with the large-scale, cyclic magnetic field. Finally, the
presence of a magnetic field leads to a significant rotational coupling between
the convection zone and underlying stably stratified core, extending deep into
the latter. Such a coupling is almost entirely absent in the HD simulation.
The high degree of (anti)symmetry about the equator
apparent on panels (c) and (d) is a true feature of these simulations, as no averaging
of hemispheres has been carried out here.

It is particularly interesting to compare Figures~\ref{flux_plots}(b) and (d) to
the corresponding diagrams presented by \inlinecite{Brun2004}, namely their
Figure~10. Note that in \inlinecite{Brun2004},
their equivalent of our integrated radial fluxes
are divided by $R^2$ and they are using CGS units instead of SI units. Moreover,
we use latitudes rather than polar angles, so that on Figures~\ref{flux_plots}(c) and (d) herein
a positive latitudinal flux implies northward transport, while
in \inlinecite{Brun2004} northward transport correspond
to negative flux values.
Their simulation differs from ours in three important ways :
i) it covers only the convecting layers; ii) it includes substantial explicit
viscosities and magnetic diffusivities throughout the simulation domain; and
iii) it does not generate a large-scale magnetic component. Even though their
mean differential rotation is qualitatively similar to ours (equatorial
acceleration, polar deceleration,
tendency towards cylindrical isocontours in equatorial regions),
major differences exist in the underlying rotational dynamics.
Viscous forces play a major role in the radial transport of angular momentum
in their simulation, teaming with Maxwell stresses to offset the Coriolis force and Reynolds
stresses throughout the whole convection zone. 
In our simulation, this dynamical balance (minus explicit
viscous force) materializes
only in the lower third of the convecting layers, with Reynolds and Maxwell stresses acting
in opposition to the Coriolis force higher up. In both simulations, the latter leads
to a net upward transport of angular momentum, and Maxwell stresses drive a
downward transport.

For the simulation analyzed on Figure~10 of \inlinecite{Brun2004},
viscous diffusion makes a lesser contribution to angular
momentum transport in the latitudinal direction
than it does in the radial direction. There
are now more similarities with our latitudinal fluxes, Reynolds and 
Maxwell stresses now opposing each other at most latitudes.
The primary difference,
besides the presence of a significant large-scale magnetic torque in our simulation,
is found with the net latitudinal angular-momentum transport by the meridional
flows, which is equatorward
at low to mid-latitudes in our simulation, but poleward
in the \inlinecite{Brun2004} simulations. In our MHD simulation, except very near
the Equator, the meridional flow and Reynolds stresses team up to sustain equatorial
acceleration. These two are resisted by magnetic forces, consistent with the Pole-to-Equator
angular velocity contrast being some three times smaller in the MHD simulation
than its purely HD parent simulation ({\it cf.} top and bottom panels on
Figure~\ref{difrot}).

The differences between these two sets of simulations most likely do not
arise exclusively from the presence of a large-scale
cyclic magnetic field in our simulations, as angular-momentum fluxes calculated
in purely HD versions of the \inlinecite{Brun2004} simulations
(see \opencite{Miesch2000} and \opencite{Brun2002}) also differ markedly from
the HD balance depicted on Figures~\ref{flux_plots}(a) and (c) herein. Looking at case (c) in 
Figure~11 from \inlinecite{Brun2002}, which is the most turbulent,
and comparing it with our Figures~\ref{flux_plots}(a) and (c), we
observe very different patterns for both the Reynolds stresses and the Coriolis
force acting on the meridional
flow. The radial fluxes show distinct depth variations,
particularly in the middle of the convection zone, but the difference is most
striking in the latitudinal fluxes distributions.
At low latitude
[$-30^{\mathrm{o}} \leq \theta \leq 30^{\mathrm{o}}$] in Figure~\ref{flux_plots}(c) and 
Figure~11(c) from \inlinecite{Brun2002}, both the Reynolds stresses and the meridional-circulation
contributions are of opposite sign when comparing both simulations. Additionally, 
there is a sign change around $\pm 30^{\mathrm{o}}$ in our simulation that has
no counterpart in theirs.
\inlinecite{Brown2008} present another set of HD simulations, all strongly turbulent
but where they also vary the rotation rate. More specifically, in their
Figure~9 they show angular-momentum fluxes for two cases:
a star which has a solar-like rotation rate in panels (a) and (b),
and one with a rotation rate five times greater than the Sun in (c) and (d). While panels (a)
and (b) remain similar to the aforecited equivalent plots in \inlinecite{Brun2002},
the more rapidly rotating case (panels (c) and (d)) reveals yet again a distinct
dynamical balance. The major players in the radial transport of angular momentum
are the Reynolds
stresses and the viscous transport terms, with the Coriolis force exerted on the meridional
circulation playing a lesser role. While latitudinal transport has a very complex profile,
the Reynolds stresses and meridional circulation terms usually
have the same sign and are counterbalanced by the viscous transport.

\begin{figure}[p!]
\begin{center}
\includegraphics[width=0.95\linewidth]{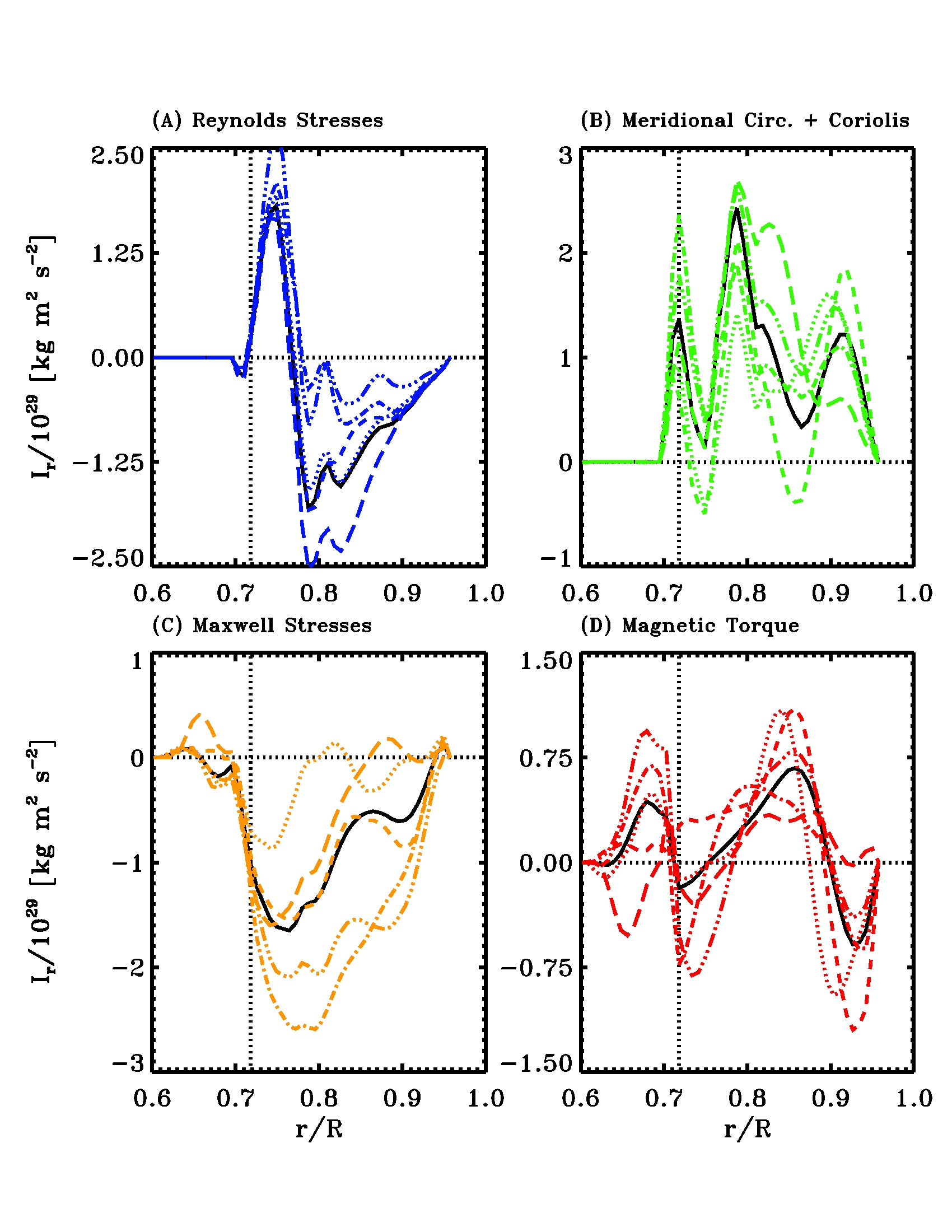}
\caption{Temporal evolution of the various radial angular-momentum flux contributions
sampled at a nine-year cadence between 54 and 90 years: 
dotted $t=54\,$years; short-dashed $t=63\,$years; dot-dashed $t=72\,$years; triple-dot-dashed
$t=81\,$years; long-dashed $t=90\,$years, indicated by the correspondingly coded vertical
line segments on Figures~\ref{tor_mag}(a) and (d).
The solid line is the corresponding
contribution to the angular-momentum transport rate
averaged over the entire simulation time, taken from
Figure~\ref{flux_plots}B.
Note the varying vertical scales on the four panels.
}\label{cycle_r}
\end{center}
\end{figure}

Another major difference lies of course with the fact that our dynamical
balance is not perfectly stationary, showing instead periodic variations
which drives the torsional oscillations visible on Figures~\ref{u_t} and \ref{u_r}.
It is therefore also interesting to examine the temporal evolution
of those angular-momentum fluxes over a magnetic half-cycle. This is carried
out on Figure~
\ref{cycle_r}, which shows the evolution of each individual flux component,
from the beginning of the second half-cycle on Figure~\ref{tor_mag}
to its end, {\it i.e.} from one minimum to the next, at a nine-year temporal cadence,
as indicated by the vertical line segments on Figures~\ref{tor_mag}(a) and (d).
Each panel also reproduces the temporal average
of the corresponding
contribution to the angular-momentum transport rate
over the full simulation duration (in black), taken
directly from Figure~\ref{flux_plots}(b).

The magnetic-torque contribution (panel (d)) shows large variations about its
temporal average, which is of course expected in view of the cyclic evolution
characterizing the large-scale magnetic field. Far less expected {\it a priori},
however, is the fact that all other flux contributions also undergo similar
variations in the bulk of the convecting layers. This is even the case for the
nominally non-magnetic contributions, namely the turbulent Reynolds stresses
and Coriolis force acting on the meridional flow. The latter's temporal variations
show little spatial coherence, whereas Reynolds stresses vary largely in unison
at all depths. A similar situation arises
with the magnetic contributions, with the magnetic torque showing a complex
spatio--temporal behavior, while temporal variations in Maxwell stresses are
very well-correlated spatially.
Also noteworthy, flux contributions associated with the small-scale flow
and magnetic field --- Reynolds and Maxwell stresses ---
fall rapidly with depth at and below the core--envelope interface,
where they show almost no temporal variability.
Consequently, the time-dependence of the rotational
coupling between the convection zone and underlying stably stratified
fluid layers is driven by the interplay between the strongly time-varying
contributions of large-scale magnetic torques and angular-momentum
advection by the meridional flow.

\begin{figure}[p!]
\begin{center}
\includegraphics[width=0.95\linewidth]{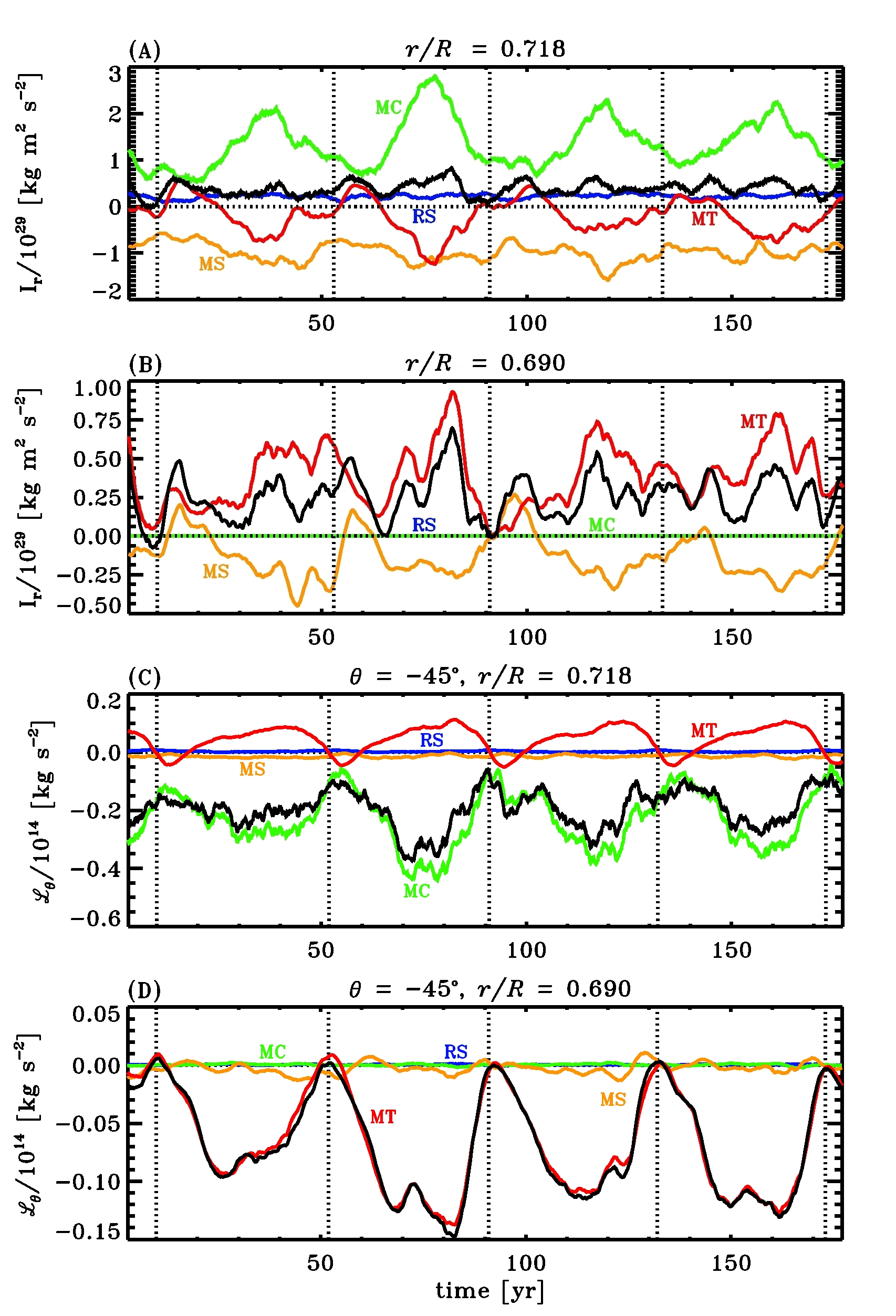}
\caption{Temporal evolutions of angular-momentum fluxes for each contribution, as labeled.
Panels (a) and (b) show the net radial angular-momentum transport rate through spherical
shells at $r/R = 0.718$ and $r/R = 0.69$. Panels (c) and (d) display time series of the
mean latitudinal flux at these same depths, both at $\theta = -45^{\mathrm{o}}$.
Note the varying vertical scales on panels (a) {\it vs.} (b), and (c) {\it vs.} (d).
The vertical dotted lines indicate epochs of polarity reversal of the
large-scale internal toroidal magnetic field, as determined from Figure~\ref{tor_mag},
equivalent here to ``solar minimum''. Here each time series has been smoothed with 
a boxcar average of width four years.
}\label{4flux_t}
\end{center}
\end{figure}

Another interesting aspect of our results relates to the rotational coupling
between the convection zone, where differential rotation is generated,
and the underlying stably stratified fluid layers.
Figure~\ref{4flux_t}(a) shows time series of the radial fluxes of angular momentum,
computed via Equation~(\ref{Ir}) separately for each of its four contributions, as labeled.
At the core--envelope interface proper ($r/R=0.718$), the net transport of angular momentum
across the corresponding spherical shell is directed upward, and it is driven primarily
by the Coriolis force acting on the meridional flow, and resisted by the magnetic
forces. Although the magnitude of
the large-scale magnetic torque
varies cyclically in phase with
the magnetic cycle, as one would have expected,
the Coriolis term does also, which results
in a net upward flux of angular momentum (in black) that does not show a well-defined
cyclic signal. This general pattern is maintained 
down to $r/R\approx 0.70$ in the stable layers, but with the disappearance
of the HD forces further below, the magnetic terms take over
completely, as shown on Figure~\ref{4flux_t}(b). The large-scale torque now drives
an upward angular-momentum flux, and is opposed by the Maxwell stresses.
Cyclic variations on the magnetic cycle frequency can still be detected,
but quasi-cyclic modulations on shorter periods are also apparent at these depths.

A similar dynamical transition as a function of depth in the stable layer
is also present in the latitudinal angular-momentum
fluxes, as illustrated on Figures~\ref{4flux_t}(c) and (d).
These time series result from the calculation of the four individual
contributions on the RHS of Equation~(\ref{Ltheta}) at latitude $-45^\circ$
and two different depths, as labeled.
At the mid-latitude core--envelope interface (panel (c)), the latitudinal flux
is directed equatorward and dominated by the meridional flow contribution, which
shows a strong cyclic signal in phase with the magnetic cycle. This is associated
with a strong driving of the meridional flow by the large-scale magnetic field
(on this point see also \opencite{Passos2012}), but with regards to the zonal dynamics,
the large-scale magnetic torque here drives angular momentum poleward, as expected
from the shearing of a latitudinally oriented poloidal large-scale magnetic field
by a latitudinal differential rotation characterized by equatorial acceleration.
Moving inward, already at $r/R=0.70$ (not shown) the magnetic torque
has reversed and
now drives angular momentum equatorward, and by $r/R=0.69$ (panel (d))
the contribution of the
meridional flow has vanished and the large-scale magnetic torque is the sole
driver of the equatorward angular-momentum flux, which remains strongly modulated
by the magnetic cycle. The Reynolds and Maxwell stresses are minor contributors
to the latitudinal flux of angular momentum at all depths in the stable layer.

The downward penetration of the large-scale magnetic fields within the stable layer allows
a rough estimate of the average numerical diffusivity characterizing the simulation
in that part of the domain. The large-scale toroidal magnetic field peaks around
$r/R=0.7$, oscillates on a $P\approx 80$-year full magnetic period, and penetrates
inward by $r/R\approx 0.05$ (see Figure~\ref{tor_mag}(d)). Equating this observed penetration
distance to the electromagnetic skin
depth $d=\sqrt{\eta P}$ leads to an estimate of the effective magnetic
diffusivity $\eta\approx 5\times 10^5\,$m$^2$ s. This is quite small, as expected
given that large velocity and magnetic field gradients, which determine the level
of implicit diffusivities within our computational framework, diminish rapidly upon
moving downward
into the stable layers. One must bear in mind, however, that the effective diffusivities
are likely much higher in the turbulent environment of the convecting layers.

\subsection{Volumetric Force Densities}

We now turn to explicit calculations of volumetric force densities, by taking
the divergence of the various contributions to the total angular-momentum flux, as appearing
on the LHS of Equation~(\ref{eq:cons}), at each grid point in the meridional $(r,\theta)$ plane.
In both HD and MHD simulations, temporally averaging each set of grid point values over four 
magnetic half-cycles again produces patterns with a high degree of symmetry with respect
to the Equator. Comparing
HD and MHD simulations reveals the most pronounced differences at mid- to high
latitudes, where even the nominally HD forces --- Reynolds stresses and
Coriolis force acting on the meridional flow --- show large differences in their spatial
distributions. This is particularly striking in the Coriolis term, which tends to
accelerate (decelerate) the zonal flow in the outer (inner) half of the convecting
layers of the HD simulation, and shows the opposite pattern in the MHD simulation.
In the MHD simulation, Maxwell and Reynolds stresses tend to oppose
each other at most locations in the meridional plane, and the two magnetic contributions
are the sole significant contributors within the underlying stable fluid layer,
consistent with Figure~\ref{flux_plots}.

\begin{figure}[h!]
\begin{center}
\includegraphics[width=0.95\linewidth]{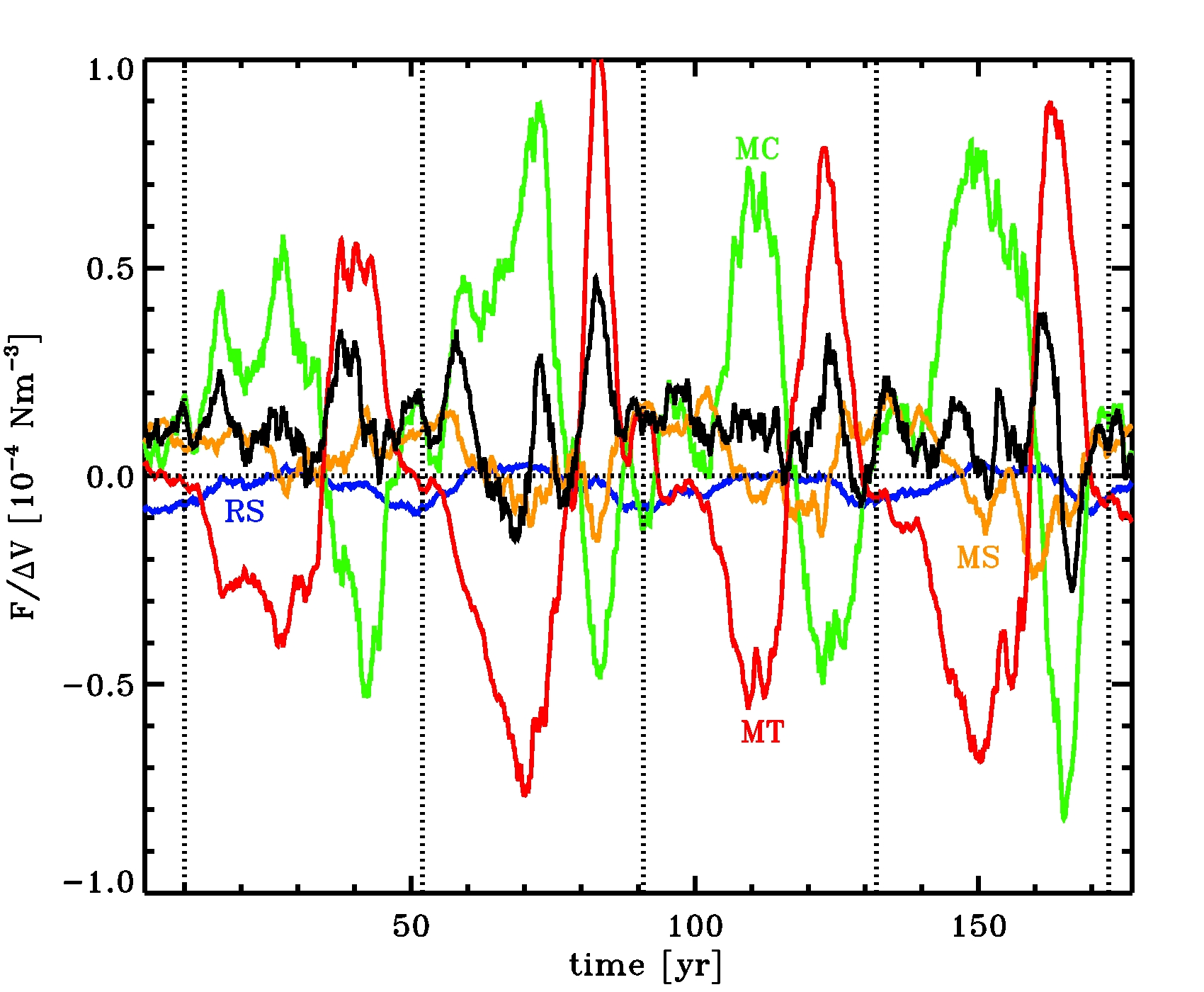}
\caption{Temporal evolution of the volumetric force density
applied on the plasma
at $r/R = 0.925$ at a latitude of $-70^{\mathrm{o}}$.
The individual force components are color-coded as:
blue -- Reynolds stresses (RS);
green -- meridional circulation with Coriolis effect (MC);
orange -- Maxwell stresses (MS);
red -- large-scale magnetic torque (MT);
black -- sum of all four contributions. The vertical
dotted line indicate times of magnetic cycle minimum, based on the zonally averaged
toroidal field at the core--envelope interface ({\it cf.}~Figure~\ref{tor_mag}).
}\label{total_forces}
\end{center}
\end{figure}

Figure~\ref{total_forces} offers a more focused look at the torsional-oscillation
dynamics at high latitudes, in the form of time series of the various azimuthal-force
components extracted at a specific grid point located at depth $r/R=0.925$ and
$-70^\circ$ latitude (southern hemisphere).
At first glance, the Reynolds and Maxwell stresses contribute very little to the zonal
dynamics at this location, which
is primarily driven by
the large-scale Lorentz force (red) and Coriolis force (green). However, these two
contributions are of similar magnitudes but 
strongly anticorrelated, and so nearly cancel each other.
The resulting
total force (black) is then of much smaller magnitude than either of these two contributions,
and comparable again to the lower amplitude Maxwell stresses (orange).

\begin{figure}[p!]
\begin{center}
\includegraphics[width=0.95\linewidth]{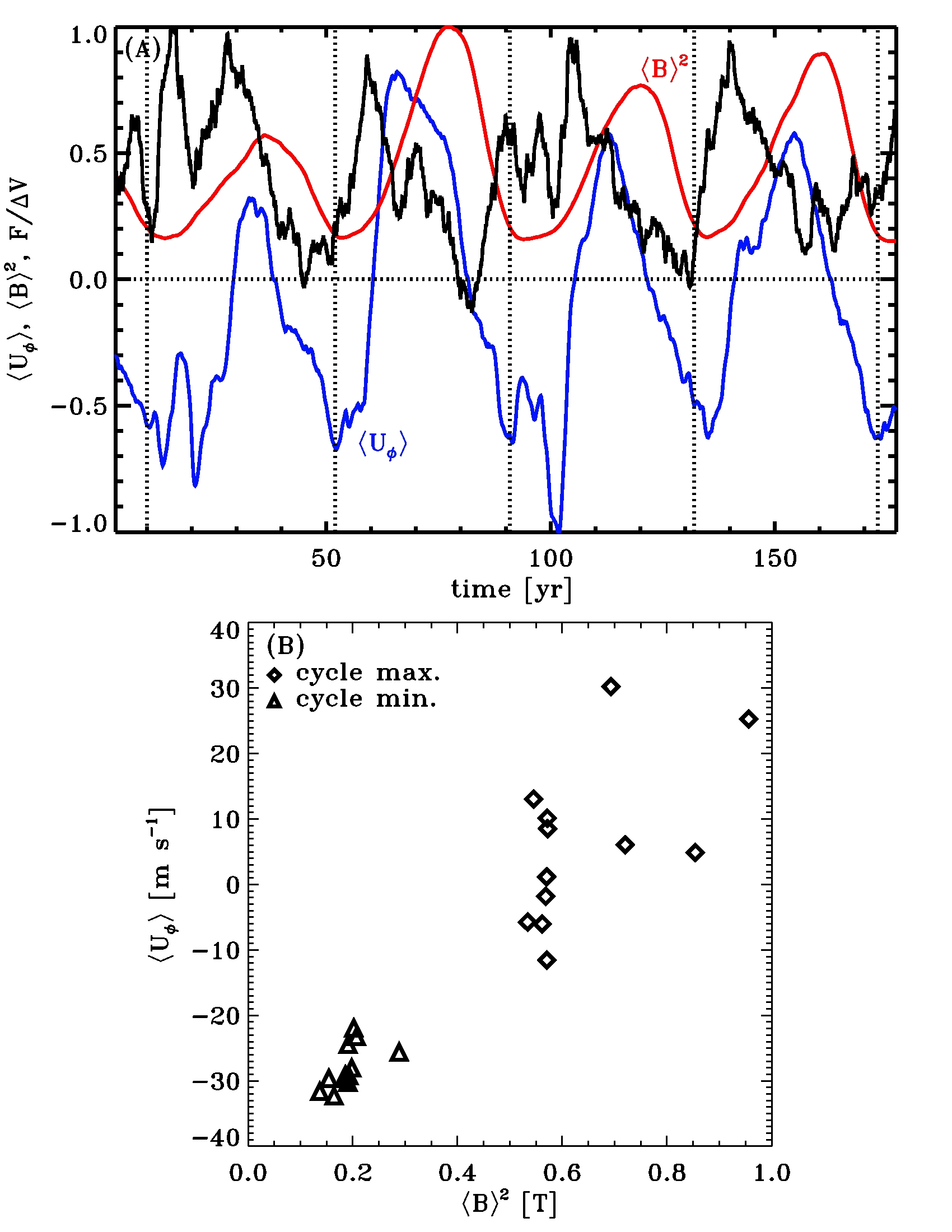}
\caption{
Panel (a) shows the temporal variations of the mean volumetric force density
(black) and mean zonal flow (blue, in the rotating frame of the simulation, thus
the negative values), both
spatially averaged over the latitudinal band $-90^\circ\leq \theta\leq -70^\circ$
and depth of the convection zone. The red line is a cycle proxy constructed by
integrating the magnetic energy over a spherical shell straddling
the core--envelope interface. All three time series have been normalized
to their peak amplitude for plotting purposes. Panel (b) shows a correlation
of magnetic energy {\it versus} torsional-oscillation amplitude
obtained by integrating the corresponding (unnormalized) time series in (a)
over a temporal window of width $\pm$ two years centered on either cycle maximum
(diamonds) or minimum (triangles), with the same operation carried out in the 
northern hemisphere high-latitude regions.
}\label{corruB}
\end{center}
\end{figure}

Figure~\ref{corruB}(a) shows time series of this volumetric force density (black)
and zonal flow speed (blue), both now averaged
radially over the convection zone ($0.718\leq r/R\leq 0.96$) and
latitudinally over a
band of angular width $\pm 20^\mathrm{o}$ centered on the South Pole.
The red line is a proxy for the cycle amplitude, constructed by
integrating the magnetic energy over a spherical shell ($0.66\leq r/R \leq 0.74$)
straddling the core--envelope interface.
All three time series have been normalized
to their peak unsigned amplitude, for plotting purposes.
Despite strong temporal fluctuations, the plot reveals that the
net zonal force peaks throughout the rising phases
of each magnetic cycle,
leading the rise in the zonal prograde flow. The latter peaks shortly before
cycle maximum, and reaches its lowest speed at times of cycle minimum
({\it cf.}~Figures~\ref{tor_mag} and \ref{u_r}).
This pattern is illustrated differently on Figure \ref{corruB}(b),
which shows a correlation plot of peak longitudinal velocity amplitude
{\it versus} magnetic energy, calculated by averaging temporally over
a $\pm$ two-year temporal window centered on either cycle maximum or minimum.
The calculation is carried out independently in each hemisphere,
and reveals a good correlation ($r \approx 0.918$).

\begin{figure}[h!]
\begin{center}
\includegraphics[width=0.95\linewidth]{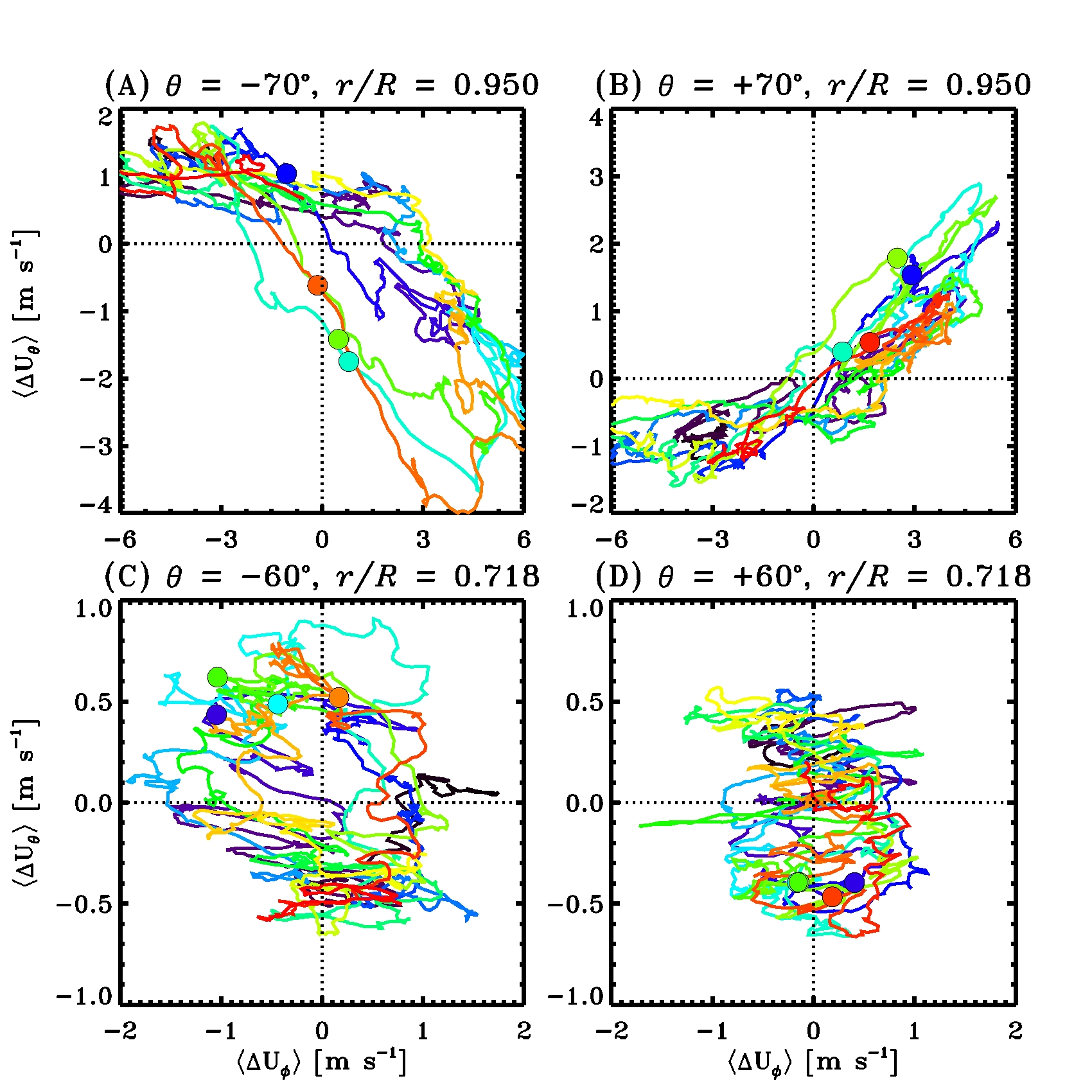}
\caption{Phase diagrams of the perturbation in latitudinal velocity plotted {\it versus}
the perturbation in longitudinal velocity for two depths and four latitudes, as
labeled above each panel.
The color sequence indicates the temporal evolution: black -- violet -- blue -- green --
yellow -- orange -- red. The colored solid dots indicate the epochs of magnetic-cycle maxima.}\label{phase}
\end{center}
\end{figure}

Figure~\ref{phase} offers yet a different look at the torsional-oscillation dynamics.
The curves are trajectories in a two-dimensional phase space defined in terms
of the zonally averaged zonal flow deviation about its temporal mean over the
simulation timespan (horizontal),
{\it versus} the zonally-averaged latitudinal
flow deviation about its own temporal mean (vertical). Four such trajectories
are shown, for meridional plane grid points located at the subsurface high latitudes
$(r/R,\theta)=(0.95,\pm 70^\circ)$ in panels (a) and (b), and mid-high-latitude
core--envelope interface
$(r/R,\theta)=(0.718,\pm 60^\circ)$ in panels (c) and (d).

In the high-latitude subsurface layers (panels (a) and (b)), both flow perturbations 
are strongly correlated in time, with the poleward latitudinal flow varying in phase
with the prograde rotational acceleration.
That local rotational acceleration (deceleration) should correlate in this manner
to the variation of the latitudinal flow component, consistent with conservation
of angular momentum in an axisymmetric fluid ring symmetrically contracting (stretching)
as it gets displaced towards (away) from the rotation axis by the latitudinal flow.
This suggests that, in this high-latitude location, the zonal dynamics are ``enslaved''
to the meridional flow dynamics. Interestingly, a similar
situation was observed by \inlinecite{Gilman1983} when analyzing the torsional oscillations
generated by his pioneering global MHD simulation of solar dynamo action, even though
magnetic driving of the meridional flow perturbation could not be firmly established.
This is all the more remarkable given that in the \inlinecite{Gilman1983} simulations,
angular-momentum transport by the meridional flow played little significant
role in sustaining the average differential rotation profile. 

Enslavement of zonal dynamics to meridional flow variations does not hold everywhere, however,
as evidenced by panels (c) and (d) of Figure~\ref{phase}. These diagrams are now constructed from
time series of zonal- and latitudinal-flow variations extracted at $\pm 60^\circ$
latitude at the
core--envelope interface. The corresponding phase diagrams of zonal- and latitudinal-flow residuals are now markedly different from their high-latitude subsurface
counterparts on panels (a) and (b), with the two flow residuals now varying cyclically
but out of phase with one another, with a phase lag $\approx\pi/2$. This indicates that
the zonal dynamics cannot be reduced to angular-momentum conservation in a
contracting (expanding) fluid ring, and results from a more complex interplay
of time-varying direct and indirect magnetically mediated forcing
(see also the companion analysis presented in \opencite{Passos2012}).

\subsection{Energetics of the Torsional Oscillations}

The conclusion drawn above can be further substantiated through an analysis
of the flux of energy to and from the various energy reservoirs defined by the
flow and magnetic field. The evolution equation for the kinetic energy
density of the flow
[$\varepsilon_F = \rho({\bf u} \cdot {\bf u})/2$]
takes here the form:
\begin{equation}
\label{EU}
\frac{\partial \varepsilon_F}{\partial t} 
= {\bf u} \cdot {\bf F}_F + {\bf u} \cdot {\bf F}_B 
+ {\bf u} \cdot {\bf F}_g ~,
\end{equation}
where the volumetric force densities associated with the flow [${\bf F}_F$],
magnetic field [${\bf F}_B$], and buoyancy [${\bf F}_g$]
have been schematically grouped on the RHS.
With ${\bf u}={\rm d}{\bf x}/{\rm d}t$,
both terms on the RHS correspond to the volumetric work done by the various forces
on, or against, the flow {\bf u}.
In what follows, we are interested in the energy flow in association
with the zonal dynamics, so we will set ${\bf u}\equiv u_\phi$,
${\bf F}_g=0$,
${\bf F}_F\equiv F_{\rm Reyn}+F_{\rm Circ}$, and
${\bf F}_F\equiv F_{\rm Maxw}+F_{\rm Magn}$, as defined by
Equations~(\ref{FReyn}) --- (\ref{FMagn}).
Table \ref{meanpow} lists the total power associated with these
zonal volumetric force
components (left column),
averaged over the full 180-year time span of the simulation segment and
integrated spatially over the southern polar caps
(middle column),
and over the full domain (right column).

\begin{center}
\begin{table}[h]
\begin{tabular}{c c c}
\hline
Powers (W) & $-90^{\mathrm{o}} < \theta < -60^{\mathrm{o}}$ & $-90^{\mathrm{o}} < \theta < 90^{\mathrm{o}}$ \\
\hline
$\int u_{\phi} F_{\mathrm{Reyn}}\,\mathrm{d}V$ & $+7.67 \times 10^{20}$ & $+2.03 \times 10^{23}$ \\
$\int u_{\phi} F_{\mathrm{Circ}}\,\mathrm{d}V$ & $+2.75 \times 10^{21}$ & $+1.88 \times 10^{23}$ \\
$\int u_{\phi} F_{\mathrm{Maxw}}\,\mathrm{d}V$ & $-3.55 \times 10^{19}$ & $-7.48 \times 10^{22}$ \\
$\int u_{\phi} F_{\mathrm{Magn}}\,\mathrm{d}V$ & $-3.12 \times 10^{21}$ & $-5.45 \times 10^{22}$ \\
\hline
\end{tabular}
\caption{Mean power obtained upon averaging in time the terms in Equation~(\ref{EU}) for two spatial
integration ranges. A positive value implies energy input into the zonal flow.} \label{meanpow}
\end{table}
\end{center}

Over the full simulation domain,
Reynolds
stresses and the Coriolis force inject energy into the zonal flow,
each contributing more or less equally,
and are resisted by both magnetic contributions, each again contributing
approximately equally to the draining of zonal kinetic energy.
Moreover, all four of these power contributions
turn out to be fairly steady in time.
This state of affairs reflects primarily
the sustenance of the mean differential rotation (Figure~\ref{difrot}).

\begin{figure}[h!]
\begin{center}
\includegraphics[width=0.95\linewidth]{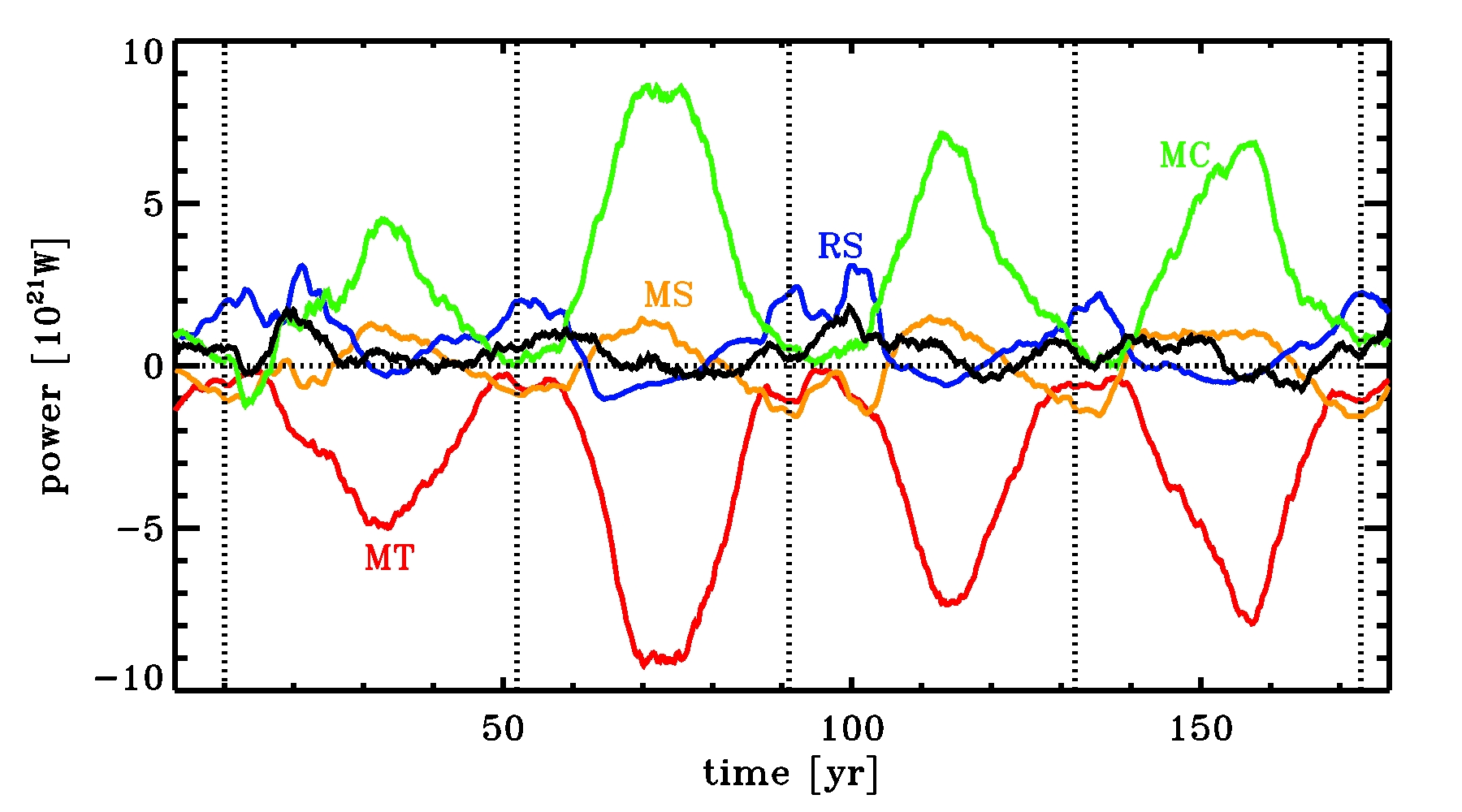}
\caption{Temporal evolution of the power of various force contributions at the South Pole of the Sun, integrated for $-90^{\mathrm{o}} < \theta < -60^{\mathrm{o}}$.
Color coding as before:
blue -- Reynolds stresses; green -- Coriolis force affecting the meridional flow;
orange -- Maxwell stresses; red -- large-scale magnetic fields; black -- total.
The vertical dotted lines delineate subsequent magnetic half-cycles.
All time series have been smoothed with a boxcar filter of width five years}\label{power}
\end{center}
\end{figure}

A different picture emerges, however,
if one focuses on high latitude regions, where torsional
oscillations have their highest amplitudes (see Figure~\ref{u_r}).
Figure~\ref{power} 
shows time series of
each power-density contribution from Equation~(\ref{EU}) integrated over a conical
volume going from the South Pole to latitude
$-60^{\mathrm{o}}$.
Energy injection into the torsional oscillations is now dominated
by the Coriolis force exerted on the
meridional circulation, with peak energy-transfer rates usually occurring
at the peak phase of each magnetic half-cycle.
Reynolds stresses still contribute significantly, but only around the times
of magnetic-polarity reversals.
The most important agent systematically extracting energy from
the zonal flow is now the large-scale magnetic
torque, doing so in a very well-defined cyclic fashion. This reflects
the action of the azimuthal Lorentz force associated with
the shearing, by the
torsional oscillations, of the strong polar magnetic fields generated
in this simulation.

The rotational energy extracted by the large-scale magnetic torque
represents an energy input into the large-scale magnetic field.
In the absence of explicit
Ohmic dissipation, magnetic energy evolves according to:
\begin{equation}
\label{EB}
\frac{\partial \varepsilon_B}{\partial t} 
= -\nabla\cdot {\bf S} - {\bf u} \cdot {\bf F}_B ~,
\end{equation}
where ${\bf S}$ is the Poynting electromagnetic energy flux, and the second
term on the RHS is the direct counterpart 
of the same term appearing on the RHS of Equation~(\ref{EU}), except of course
for the sign.
Torsional oscillations, far from being directly driven
by large-scale magnetic torques, instead divert energy back into
the large-scale magnetic field through the agency of magnetic torques.
The conclusion is that direct magnetic driving of torsional oscillations
does not represent a saturation mechanism for
the global dynamo operating in this simulation. However, 
the magnetic cycle
does drive large fluctuations in the meridional flow, and the Coriolis force
acting on this cyclically forced flow turns out to
be the main driver of torsional oscillations.
Indeed, the energy analysis presented in Section 5 of \inlinecite{Passos2012} indicates
that magnetic driving of the latitudinal flow is the primary sink of magnetic
energy in this simulation. Schematically,
the energy flow is thus of the form:\medskip

\noindent Magnetic energy $\to$ meridional flow $\to$ torsional oscillations
$\to$ (numerical) dissipation.

\medskip
The most important ``take-home'' message of the above analyses is
the following: torsional oscillations are not driven by a cyclic,
magnetically mediated perturbation superimposing itself on
an otherwise steady hydrodynamical zonal
balance. Instead, all angular-momentum flux contributions, including those of a
purely HD nature, are strongly modulated by the magnetic
cycle. In these simulations, torsional oscillations
are a fully nonlinear and truly MHD phenomenon.

\section{Concluding Remarks}

In this article, we have carried out a focused analysis of one of the 
implicit large-eddy MHD simulations of solar convection
computed by \inlinecite{Ghizaru2010} (see also \opencite{Racine2011}).
To the best of our knowledge,
these simulations remain unique in generating a spatially well-organized
large-scale magnetic-field component undergoing regular polarity reversals
in a manner resembling in many ways the solar magnetic cycle. We have shown
that a well-defined rotational torsional oscillation signal is present in
the simulation, showing a high degree of similarity with
those observed in the Sun, including: i) frequency
twice that of the magnetic cycle; ii) greater amplitudes in polar
and subsurface regions, peaking at a few nHz; iii) peak prograde phase
coinciding approximately with the peak in large-scale magnetic field;
iv) diverging double-branch latitudinal structure; v) depth-independent phase
at most latitudes.

We investigated rotational dynamics by first
computing from the simulation output the various contributions to angular
momentum fluxes, in the MHD simulation as well as a parent, unmagnetized
simulation otherwise operating under the same numerical and physical
parameter settings. We could show that in the MHD simulation, the presence
of a large-scale cyclic magnetic field drives torsional oscillations
not just directly through the associated large-scale magnetic torque, but
also indirectly by modulating the other forces influencing zonal dynamics, most
notably the transport of angular momentum by meridional flow. In fact,
{\it all} force components driving the zonal flows undergo cyclic variations
driven by the magnetic cycle, including the nominally ``small-scale''
Reynolds and Maxwell stresses.

We also examined the dynamical character of the rotational coupling
between the convecting layers and underlying stably stratified fluid layers.
Because of the low-dissipative properties of the numerical scheme underlying
the simulations, a tachocline-like shear layer builds up immediately beneath
the nominal base of the convective layers, within which significant
radial and latitudinal cyclically-varing fluxes of angular momentum
develop. The net radial flux is determined by competition between
the meridional flow and Maxwell stresses within the tachocline
($0.7 \leq r/R\leq 0.718$) and by the magnetic torque and Maxwell stresses
below. The latitudinal flux shows a strong cyclic signal,
in phase with the
magnetic cycle. It is dominated by the meridional flow within the
tachocline, and by the large-scale magnetic torque below.
The upper part of the stable layer is here an important player in
setting the global cycle of angular-momentum redistribution --- and thus torsional
oscillations --- within the convection zone 
(on these issues see also \opencite{Gilman1989}).

Turning to a simple analysis of the energetics of torsional oscillations,
we could also show that the primary direct power source for torsional
oscillations arising in the simulation is the action of the Coriolis force
on large-scale meridional fluid motion, with large-scale magnetic torques
acting to oppose these oscillations at most phases of the cycle.
This is in agreement with a parallel investigation by \inlinecite{Passos2012}, who
carried out a similar energy analysis for the meridional flow and could show
that magnetic driving of this flow represented the primary sink of magnetic
energy. This suggests that saturation of global dynamo action in this simulation
occurs through the magnetic driving of flows on large spatial scales, magnetic
energy being first diverted into the meridional components, and subsequently,
through angular-momentum conservation, into torsional oscillations, where it
eventually damps through turbulent stresses.

This state of affairs, should it carry over to the real Sun, has interesting
consequences with regards to attempts to use fluctuations in large-scale
flows as precursors to the solar-cycle amplitude. More specifically, taken jointly
with the results presented by \inlinecite{Passos2012}, our analysis suggests that
fluctuations in the meridional flow may be better potential precursors
than torsional oscillations, because the bulk of magnetic driving of large-scale
flows occurs on and through this flow component.
Re-analysis of extant surface Doppler measurements 
back to the beginning of Cycle 22 \cite{Dikpati2010,Ulrich2010}
have indeed revealed significant cycle-to-cycle variations in the surface
meridional flow.
We are currently pushing our simulations
much further in time, as well as under different parameter regimes,
which should 
allow a statistically sound investigation of the precursor
potential of torsional oscillations and surface latitudinal flow
variations.

\medskip

Acknowledgements : We are indebted to an anonymous referee 
for useful comments and suggestions.
This work was supported by Canada's Natural Sciences and Engineering Research
Council, Research Chair Program, and by the Canadian Space Agency's Space
Science Enhancement Program (grant \# 9SCIGRA-21).
PB is also supported in part through a graduate fellowship from the Universit\'e
de Montr\'eal's Physics department. PKS acknowledges partial support by 
the NSF grant OCI-0904599. NCAR is
sponsored by the National Science Foundation.

     % format of references provided by the journal (.bst)
\bibliographystyle{spr-mp-sola}
%\bibliographystyle{spr-mp-sola-cnd} %% Alternative style: no title,
                                      % no concluding page. 

     % name your Bibtex file containing your references (.bib)
\bibliography{bib_art2}  
 
\end{article}

\end{document}